\documentclass[fleqn,usenatbib]{mnras}

\usepackage{newtxtext,newtxmath}
\usepackage{appendix}
\usepackage{comment}
\usepackage{xcolor}

\usepackage[T1]{fontenc}

\DeclareRobustCommand{\VAN}[3]{#2}
\let\VANthebibliography\thebibliography
\def\thebibliography{\DeclareRobustCommand{\VAN}[3]{##3}\VANthebibliography}

\usepackage{graphicx}	
\usepackage{amsmath}	

\title[Orbital properties of intracluster stars]{Intracluster light is a biased tracer of the dark matter distribution in clusters}

\author[J. Butler et. al.]{
J. Butler,$^{1}$\thanks{E-mail: joseph.butler@nottingham.ac.uk}
G. Martin,$^{1}$
N.~A.~Hatch,$^{1}$
F. Pearce,$^{1}$
S. Brough$^{2}$
and Y. Dubois$^{3}$
\\
$^{1}$School of Physics and Astronomy, University of Nottingham, University Park, Nottingham NG7 2RD, UK\\
$^{2}$School of Physics, University of New South Wales, NSW 2052, Australia\\
$^{3}$Institut d’Astrophysique de Paris, CNRS, 98 bis blvd Arago, F-75014 Paris, France
}

\date{Accepted 2025 April 7. Received 2025 April 4; in original form 2025 February 20}

\pubyear{2025}

\begin{document}
\label{firstpage}
\pagerange{\pageref{firstpage}--\pageref{lastpage}}
\maketitle

\begin{abstract}
The diffuse stellar component of galaxy clusters known as intracluster light (ICL) has been proposed as an observable tracer of the cluster's dark matter (DM) halo. Assessing its reliability as a DM tracer requires understanding how the intracluster stars are energetically linked to the underlying DM distribution, which we investigate at $z\approx0$ in 12 galaxy clusters with $M_{178} = 1.18 - 3.71 \times 10^{14}\,\textrm{M}_\odot$ from the {\sc Horizon-AGN} simulation. We quantify the orbital energies of these components by their mean specific energies ${\langle \varepsilon \rangle}$, and find that this quantity is $\approx$\,25\,per cent lower for the intracluster stars than the DM, whilst the energetics of the satellite galaxies (a standard DM tracer) are only marginally ($\approx$\,5\,per cent) higher than the DM. Importantly, the lower ${\langle \varepsilon \rangle}$ of the intracluster stars compared to the DM is robust against the precise separation between the brightest cluster galaxy (BCG) and the ICL. The specific energy distribution of ICL stars is concentrated towards lower energies and poorly samples the higher energies, where much of the DM resides. Consequently, the intracluster stars have velocity distributions with lower typical speeds and a more centrally-concentrated density profile than the DM. We also find that intracluster stars have more radially-biased orbits than the DM, indicating these components have distinct orbital distributions. This study demonstrates that although the morphology of the ICL may match the DM halo, the ICL is a biased tracer of DM, and these biases must be understood in order to infer properties of the DM from the ICL.
\end{abstract}

\begin{keywords}
methods: numerical -- galaxies: clusters: general -- galaxies: interactions -- galaxies: kinematics and dynamics
\end{keywords}



\section{Introduction}

Galaxy clusters are the most massive collapsed structures in the Universe, with their mass distribution dominated by their dark matter (DM) halo. These haloes initially form via the gravitational collapse of the highest peaks of primordial density fluctuations \citep{Zeldovich_1970}, and then grow hierarchically via accretion of less massive DM haloes and the direct accretion of DM from less dense regions. Independent measures of cluster masses via gravitational lensing \citep[e.g.][]{Grossman_1989}, observations of X-rays emitted by the hot intracluster medium \cite[e.g.][]{Eyles_1991}, the Sunyaev-Zeldovich effect \citep{Sunyaev_Zeldovich_1972} and galaxy velocity dispersions \citep{Zwicky_1933} imply that DM must account for 80 to 90 per cent of the mass of these structures (e.g. \citealt{Gonzalez_2007, Gonzalez_2013}).

Beyond their integrated properties, the shape of DM haloes offer a probe of the nature of DM. Different models of DM (see \citealt{Feng_2010} for a review) such as self-interacting DM \citep[e.g.][]{Spergel_2000, Elbert_2015, Tulin_2018} and warm DM \citep[e.g.][]{Lovell_2012, Lovell_2014} provide possible non-baryonic solutions to small-scale challenges to standard cold DM, such as the core-cusp problem \citep{Moore_1994, Blok_2010}. Accurate measurements of the DM mass distribution are therefore needed to constrain these solutions. Since DM is non-luminous by definition, measurements of the structure, shape and morphology of cluster haloes must utilise visible tracers of the gravitational potential it generates. One approach is measuring the gravitational lensing of background galaxies, which in turn can be used to reconstruct the mass distribution of the galaxy cluster \citep[see for reviews:][]{Kneib_2011, Hoekstra_2013}. Alternatively, visible (baryonic) cluster components can be employed as tracers of the gravitational potential; typical tracers include the dynamics of satellite galaxies \citep[e.g.][]{Gifford_2013} and the hot intracluster medium \citep[e.g.][]{Borgani_2001, Ettori_2013}, the latter of which can be detected in X-rays emitted via thermal bremsstrahlung. However, all of these methods have limitations. Gravitational lensing maps can be limited by systematics such as foreground and background contamination, misidentified multiple objects or a need for stacking to enhance weak signals. Dynamical mass measurements are limited by the number of satellite galaxies in an individual cluster, resulting in a noisy signal or else requiring multiple clusters to be stacked. Finally, the intracluster medium is collisional, meaning its distribution can differ greatly from that of the DM halo \citep{Smith_2016a}, especially in unrelaxed systems \citep[e.g. the Bullet Cluster,][]{Clowe_2004, Markevitch_2004}.

In recent years, the intracluster light (ICL) has been postulated as another possible tracer of the DM halo in galaxy clusters. This light is emitted from a diffuse stellar component consisting of stars gravitationally bound to the cluster but not to any particular galaxy (see \citealt{Montes_2022} for a review). While the existence of ICL has been known for a long time \citep{Zwicky_1951}, its low-surface-brightness nature \citep[e.g. $\mu_{r}(3\sigma, 10^{\prime\prime}\times10^{\prime\prime})>26\,\textrm{mag}\,\textrm{arcsec}^{-2}\,$,][]{Montes_2021, Martínez-Lombilla_2023, Brough_2024} has prevented large statistical studies of the resolved properties of the ICL. This has begun to change in the past decade, with the Dark Energy Survey \citep[DES;][]{DES_2005} e.g. \citet{Golden-Marx_2023, Golden-Marx_2025} and the Hyper-Suprime Cam Subaru Strategic Program \citep[HSC-SSP;][]{Aihara_2018, Aihara_2022}. The advent of the Vera C. Rubin Observatory’s Legacy Survey of Space and Time \citep[LSST;][]{Ivezic_2019} and the European Space Agency’s {\it Euclid} Wide Survey \citep[EWS;][]{Scaramella_2022} will deliver systematic deep and wide imaging of the ICL across a range of redshifts, cluster masses and dynamical states. 

The primary formation mechanisms of the ICL are mergers between satellite galaxies and the brightest cluster galaxy \citep[BCG;][]{Willman_2004, Murante_2007, Conroy_2007}, and the tidal stripping of satellite galaxies \citep[][]{Rudick_2009, Contini_2014, Brown_2024, Golden-Marx_2025}. For very massive clusters ($M_{178}>10^{15}\,\textrm{M}_\odot$), preprocessing is thought to be an important secondary mechanism, where the intracluster or intragroup stars of infalling haloes are accreted \citep{Mihos_2004, Rudick_2006, Contini_2014, Cañas_2020}. The build-up of the ICL is therefore linked to the hierarchical assembly of galaxy clusters \citep{Golden-Marx_2025}. Since much of the DM accreted onto clusters is also accreted from infalling galaxy and group haloes, the orbital properties of the intracluster stars should be related to those of the DM. At the stellar densities that produce ICL, both intracluster stars and DM are collisionless particles, i.e. they do not share energy via collisions between themselves or other cluster components, and thus their subsequent dynamics are entirely determined by the evolving potential of the cluster. Since DM is the dominant matter component (with the exception of within the central regions of galaxies) the gravitational potential almost entirely reflects the spatial distribution of DM. Therefore, the ICL (emitted by intracluster stars governed by this potential) has promise as a luminous tracer of the DM halo.

Observational studies have found a strong connection between ICL and the host DM halo. \cite{Kluge_2021} reported positive correlations between BCG + ICL brightness and cluster properties such as mass, radius and velocity dispersion, indicating that both the BCG and the ICL correlate with the host cluster properties, which provides further motivation to investigate the ICL as a possible DM tracer. Furthermore, the morphologies of the ICL and DM in clusters has been compared by \cite{Montes_Trujillo_2019}. They showed that the shape of the ICL contours are well-matched to the total mass contours, in the central regions of the Hubble Frontier Field clusters, better than that of X-ray contours. This finding has also been replicated in hydrodynamical simulations \citep{Alonso_Asensio_2020, Yoo_2024}. Although the morphologies of the ICL and DM appear to match, many simulations have shown that the radial density profiles do not. The ICL tends to be more centrally concentrated, such that its radial density profile is much steeper than the DM \citep{Alonso_Asensio_2020, Contreras_Santos_2024}. 

The state of the cluster is described by its six-dimensional phase space, but can be more conveniently described in terms of orbits by transforming to orbital energy and angular momentum. If the distributions of the orbital energies and orbital angular momenta of the ICL stars matches the DM well, then the ICL could be used as an unbiased tracer of the DM. However, if these distributions are significantly different, then these differences must be quantified and their physical origin understood in order to safely infer properties of the DM from the ICL.

In this paper we investigate the orbital energetics and orbital angular momenta -- as inferred by the orbital anisotropy -- of the intracluster stars and DM, using 12 clusters with total masses $1.18\,-\,3.71\,\times10^{14}\,\textrm{M}_{\odot}$ at ${z=0.06}$ from the {\sc Horizon-AGN} cosmological hydrodynamic simulation.

This paper is organised as follows. In Section\,\ref{sec:simulation}, we outline the details of the simulation and selection of the cluster sample. In Section\,\ref{sec:method}, we define the BCG, ICL and satellite galaxies, describe the calculation of liberation time for the intracluster stars, and define the specific energies and orbital anisotropies of the cluster components. In Section\,\ref{sec:results}, we present our results, showing that the intracluster stars generally have lower specific energies and a more radially biased orbital anisotropy than both the satellite galaxies and DM. In Section\,\ref{sec:discussion}, we discuss potential reasons why the orbital properties of the intracluster stars are different to the satellite galaxies and the DM, and the observable implications of these differences. We conclude in Section\,\ref{sec:conclusions} by summarising our main results.

\section{Simulation}\label{sec:simulation}

In this work, we utilise the cosmological hydrodynamic simulation, {\sc Horizon-AGN} \citep{Dubois_2014}. In brief, a standard $\Lambda$CDM cosmology is adopted with the following parameters: $\Omega_\textrm{m}= 0.272$, $\Omega_\Lambda= 0.728$, $\Omega_\textrm{baryon}= 0.045$, $\sigma_8 = 0.81$, $H_\textrm{0} = 70.4\,\textrm{km}\,\textrm{s}^{-1}\,\textrm{Mpc}^{-1}$ and $n_\textrm{s}=0.967$, which are compatible with the \textit{WMAP7} cosmology \citep{Komatsu_2011}, and also with \cite{Planck_2014} within a 10 per cent relative variation. The simulation box with side length $100\,h^{-1}\,\textrm{comoving Mpc}$ contains $1024^3$ DM particles with a mass resolution of $M_\textrm{DM,res} = 8\times10^7\,\textrm{M}_\odot$, and an initial gas resolution of $M_\textrm{gas,res} = 1\times10^7\,\textrm{M}_\odot$. The adaptive mesh refinement code {\sc ramses} \citep{Teyssier_2002} is used as the N-body and hydrodynamical solver, which has up to 7 levels of refinement according to a quasi-Lagrangian criterion down to a spatial resolution of $\Delta x = 1\,\textrm{kpc}$. An extra level of refinement is added every time the expansion scale factor $a$ doubles in order to keep the minimum cell size approximately constant in proper units. Gas heating from a uniform UV background takes place from $z=10$ \citep[following][]{Haardt_1996} and gas cooling occurs via H, He and metals \citep[following][]{Sutherland_1993}. Details of the implementation of star formation, stellar feedback and active galactic nuclei (AGN) feedback can be found in \cite{Kaviraj_2017}.

Cosmological simulations must make a compromise between box size (thus sample size) and mass resolution. Given the significant impact of resolution on tidal stripping \citep{Martin_2024}, we prioritise resolution in our choice of simulation to study the resolved properties of the ICL. {\sc Horizon-AGN} has high stellar mass ($M_\textrm{star,res} \approx 2\times10^6\,\textrm{M}_\odot$) and spatial resolution, comparable to other cosmological simulations (e.g. TNG100, \citealt{Pillepich_2018}) and cluster zoom-in simulations (e.g. Hydrangea, \citealt{Bahé_2017}) with similar sample sizes in our chosen cluster mass regime. Additionally, many studies have used {\sc Horizon-AGN} to investigate the ICL \citep{Cañas_2020, Brough_2024, Brown_2024, Kimmig_2025}. 

Apart from choosing black hole (BH) feedback parameters to match observed BH mass relations at $z=0$, {\sc Horizon-AGN} is not otherwise calibrated to the local Universe. \cite{Kaviraj_2017} find that the aggregate star formation histories of the model galaxies in {\sc Horizon-AGN} broadly reproduce that of galaxies in the real Universe, such that important bulk properties (luminosity functions, stellar mass functions, the star formation main sequence, the cosmic star formation history and rest-frame ultraviolet--optical--near infrared colours) are in good agreement with observational data in the redshift range $0<z<6$.

The simulation begins at $z\approx38$, with the initial conditions generated using {\sc mpgrafic} \citep{Prunet_2008}. The particle data is saved with a time resolution of $\approx 20\,\textrm{Myr}$, resulting in 782 fine snapshots up to $z=0.018$. In each snapshot, DM haloes are identified using the {\sc AdaptaHOP} structure finder \citep{Aubert_2004} applied to the DM particles. Specifically, the updated version \citep{Tweed_2009} is used which ensures substructures always have lower mass then their host (sub)haloes. First, the local density associated with each particle is computed using the 20 nearest neighbours. Local maxima are identified by walking through the density field. Particles with a density above the threshold of 80 times the total matter density $\rho_\textrm{M}$ of the entire simulation box are then linked with their closest local maxima. Groups of particles containing at least 100 particles are then defined as (sub)structures, resulting in a minimum detectable DM (sub)halo mass of $8\times10^9\,\textrm{M}_\odot$. Galaxies are identified in a similar way independently from the DM haloes, with the original {\sc AdaptaHOP} structure finder applied to the star particles. However, the density threshold is instead $178\,\rho_\textrm{M}$, and the minimum number of particles in a (sub)structure is 50. This corresponds to a minimum detectable galaxy mass of $\sim\,10^8\,\textrm{M}_\odot$. Galaxies are then associated with DM haloes by linking the most massive galaxy within $0.1\,R_\textrm{178}$ of each DM halo, where $R_\textrm{178}$ is the radius of a sphere around the halo centre with average DM density $\bar{\rho}_\textrm{DM} = 178\,\rho_\textrm{M}$.

We generate merger trees from these fine snapshots using the {\sc TreeMaker} algorithm \citep[originally developed by][]{Hatton_2003}. This scheme identifies descendants of a structure by the maximum mass overlap between the structure in a given snapshot and structures in the subsequent snapshot, and similarly identifies main progenitors of a structure by the maximum mass overlap between the structure in a given snapshot and structures in the previous snapshot. The updated version \citep{Tweed_2009} expands this scheme to include substructures in the merger trees. From these 782 fine snapshots, we select 12 at intervals of $\sim1\,\textrm{Gyr}$ from $z=2.871$ to $z=0.056$; for the remainder of this paper, the term ‘snapshots’ will refer to these 12 coarse snapshots. We chose the 14 most massive DM haloes from this final snapshot ($z=0.06$) as cluster haloes, with a $M_{178}$ range of $1.18 - 8.30 \times 10^{14}\,\textrm{M}_\odot$, where $M_{178}$ is the total (inclusive) mass within $R_{178}$. The two most massive clusters are undergoing major mergers, and thus their DM and ICL distribution have multiple spatially-distinct peaks. Much of the later analysis will involve evolution with BCG--centric radial distance and assumes both a level of spherical symmetry and that the system is approximately virialised -- these assumptions are clearly broken for these two clusters, and thus they are removed from the sample.

The remaining 12 clusters, with a $M_{178}$ range of $1.18 - 3.71 \times 10^{14}\,\textrm{M}_\odot$, are chosen as our cluster sample. The properties of these clusters are displayed in Table\,\ref{tab:cluster_properties}. The galaxy considered to be the BCG of each cluster is the most massive galaxy within $0.1\,R_{178}$ of the halo centre in the final snapshot, and the main progenitor of this galaxy is considered to be the BCG in prior snapshots. For each cluster we extract cubes with side lengths of 8\,Mpc centred on the BCG for all 12 snapshots.

\section{Method}\label{sec:method}
\subsection{Defining the BCG, ICL \& satellite galaxies}\label{sec:defining_bcg_icl}

\begin{figure*}
	\includegraphics[width=\textwidth]{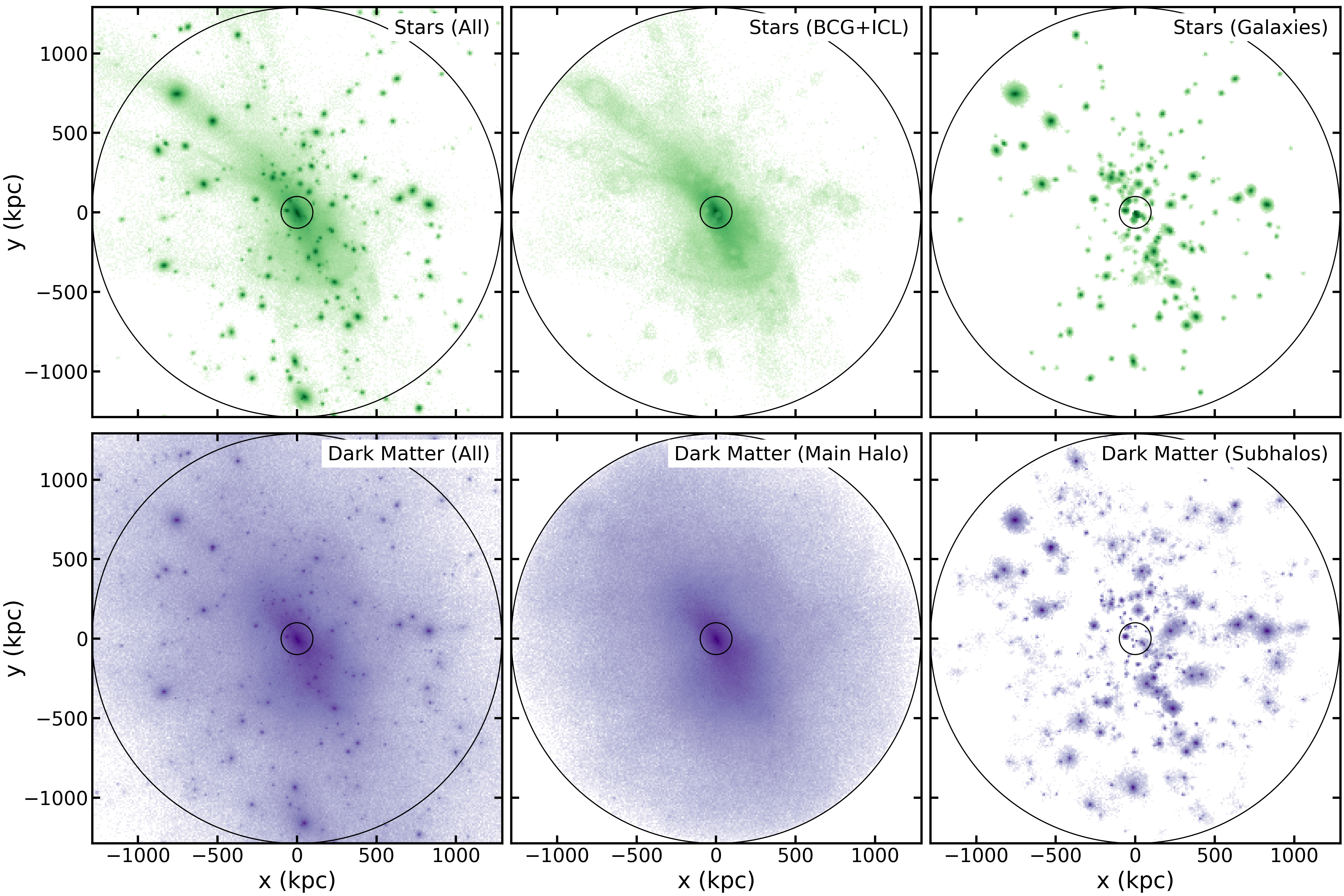}
    \caption{Projected surface density map of the stars (top, green) and DM (bottom, purple) along two arbitrary spatial axes centred on the BCG for Hrz001 at ${z\approx0}$. In all panels, darker colours represent higher projected surface densities on a logarithmic scale. The panels show: all star particles in a cube of volume $(R_{178})^3$ (top left); the BCG + ICL (top middle); the satellite galaxies (top right); all DM particles in a cube of volume $(R_{178})^3$ (bottom left); the main halo (bottom middle); and the subhaloes (bottom right). The inner and outer black circles indicate 100\,kpc and $R_{178}$ around the centre respectively.}
    \label{fig:stars_dm_map}
\end{figure*}

We investigate the orbital energies and orbital angular momenta of the intracluster stars and dark matter. There is no straightforward way to separate the ICL from the BCG, both in simulations and observations (\citealt[][]{Cui_2014, Pillepich_2014, Jiménez-Teja_2016, Cañas_2019, Montes_2022, Canepa_2025}; see \citealt{Brough_2024} for a recent comparison). Thus we begin by extracting the combined BCG and ICL component, along with the main DM halo for each cluster.

To achieve this, substructures (galaxies in the stellar component and subhaloes in the DM component) must be isolated. While {\sc AdaptaHOP} does identify substructures, it does not use an unbinding procedure to determine which particles are bound to the substructures, and therefore may attribute some loosely-bound star particles on the outskirts of galaxies to the BCG + ICL component. Moreover, even structure finders that incorporate unbinding procedures or leverage full 6-D phase-space information can yield differing results \citep{Knebe2013}. 

Since the orbits of those star particles just outside of the {\sc AdaptaHOP} structure boundary may still follow the local potential well generated by the associated satellite galaxy instead of the global cluster potential well, which reflects the global DM distribution, we impose a more stringent cut on the star particles that make up the BCG + ICL. We do this by imposing an additional spherical cut around the centre of each satellite of radius $4\,r_\textrm{sat}$, and attribute all star particles in this region to the satellite. For $M_\textrm{stellar} < 10^{11}\,\textrm{M}_\odot$, $r_\textrm{sat}$ is a constant $r_\textrm{const}$, set as the $84^\textrm{th}$ percentile of the satellite effective radii, $r_\textrm{eff}$, in this $M_\textrm{stellar}$ range within all clusters in the given snapshot, where $r_\textrm{eff}$ is the half-stellar mass radius. For $M_\textrm{stellar} \geq 10^{11}\,\textrm{M}_\odot$, $r_\textrm{sat}=r_\textrm{eff}$ if $r_\textrm{eff} > r_\textrm{const}$, otherwise $r_\textrm{sat}=r_\textrm{const}$.

These choices are informed by the following factors. First, there are orphan galaxies within the simulation. This justifies using a quantity related to the stellar $r_\textrm{eff}$ rather than the subhalo $R_\textrm{178}$. Second, we find that the $r_\textrm{eff}$ distribution at a given redshift is approximately constant\footnote{This is primarily a numerical effect due to gravitational force softening at the spatial resolution limit of the simulation.} with stellar mass up to $M_\textrm{stellar} \approx 10^{11}\,\textrm{M}_\odot$, but with some significant outliers. Third, the substructure mass recovery fraction of most structure finders decreases towards the centre of clusters \citep{Muldrew_2011}, so {\sc AdaptaHOP} may underestimate $r_\textrm{eff}$ for satellites near the cluster centre (although this is less of an issue for stars than DM). These two prior reasons, along with the aim to minimize contamination of the ICL from galactic outskirts, justify using a constant value $r_\textrm{const}$ that is related to the upper end of the $r_\textrm{eff}$ distribution. Finally, although the factor of 4 in $4\,r_\textrm{sat}$ is arbitrary, \cite{Ahvazi_2024a} also impose a similar cleaning procedure on the satellites galaxies using a factor of 4 times $r_\textrm{eff}$. We test factors of 6, 8 and 10 and find our results involving the ICL are robust to this choice. Further details of these tests are given in Section\,\ref{sec:energetics}.

The BCG + ICL for each cluster is defined as any star particle within $R_\textrm{178}$ from the centre of the BCG that is not attributed to another galaxy. The main DM halo is defined in a similar way, with all DM particles attributed to subhaloes or beyond $R_\textrm{178}$ removed, but centred on the BCG rather than the centre of the DM halo itself. An example can be seen in the projected surface density map of the cluster Hrz001 at $z\approx0$ (Fig.\,\ref{fig:stars_dm_map}), where the middle panels show the BCG + ICL (top) and main DM halo (bottom), and the rightmost panels show the removed galaxies (top) and subhaloes (bottom).

To separate the intracluster stars from those that belong to the BCG at $z\approx0$, a spherical aperture of radius 100\,kpc is placed around the centre of the BCG, and all star particles outside this radius are considered intracluster stars. We define the spherical shell with inner radius 100\,kpc and outer radius $R_{178}$ as the ‘ICL region’. This aperture is a relatively conservative radial cut for the ICL -- for example, \cite{Brough_2024} use 30, 50 and 100\,kpc cuts between the BCG and ICL. The transition radius, where the density profile of the ICL becomes dominant over the BCG, is typically identified to be between 30\,kpc and 100\,kpc \citep{Zhang_2019, Montes_2021, Golden-Marx_2025, Kluge_2024, Zhang_2024}. \cite{Proctor_2023} identify a transition radius for $10^{14}\,\textrm{M}_{\odot}$ haloes of $\sim100\,\textrm{kpc}$, matching well with our BCG--ICL cut.

We discuss how different BCG--ICL radial separations affect our results in Section\,\ref{sec:energetics}.

\subsection{Liberation time of the intracluster stars}\label{sec:liberation_time}

This work will not be concerned with the progenitor objects of the ICL (see \citealt{Brown_2024} for this analysis on a similar sample of {\sc Horizon-AGN} clusters), however, the time at which a star particle becomes part of the ICL, the so-called liberation time, is relevant. Since the ICL is assembled from many individual stripping and merging events, the orbital energy and angular momentum of the progenitor galaxies and the cluster itself at the time of liberation are imprinted onto the phase space distribution of each of these ICL subpopulations. Although these individual ICL subpopulations will evolve with time, the particles are collisionless so only phase mixing will occur. Once a star particle is removed from a galaxy, its orbital energy should only be impacted by the large-scale potential, so trends in the ${z\approx0}$ ICL properties with liberation time can reveal how the galaxy or cluster properties at the time of liberation impact the present ICL properties. 

The $z\approx0$ intracluster stellar particles are assigned a liberation time based on the following criteria:

\begin{enumerate}
    \item The particle's first appearance in the cutout cube must be in a satellite.
    \item The particle must undergo a transition from a satellite to the BCG + ICL (a ‘liberation event’).
    \item If a particle undergoes multiple liberation events, the first instance is selected as the liberation time.
\end{enumerate}
In extreme scenarios the main progenitor branch of the merger trees may break \citep{Srisawat_2013}, which we find to occur for clusters Hrz078, Hrz137 and Hrz157 in our sample, so we remove them for the liberation time analysis.

For this part of the method, the definition of a satellite in every snapshot is extended to encompass a sphere of radius $8\,r_\textrm{sat}$. A factor of 8 is used here instead of the factor of 4 used to define galaxies in Section\,\ref{sec:defining_bcg_icl}, to remove spurious liberation events due to particles with highly radial orbits continually passing between the inside and outside of the galaxy border. It is emphasised that the term ‘liberation’ is used here not to imply a binding energy criterion has been applied, but instead as a neutral term that describes particles that have been removed from their satellites without attempting to identify the process by which this happened (e.g. tidal stripping or satellite-BCG mergers). A particle may therefore appear to undergo multiple liberation events if it overlaps spatially with another satellite in a subsequent snapshot, and so the first instance is selected as the liberation time.

The fraction of $z\approx0$ ICL particles assigned a liberation time averaged between the 9 clusters is ${f_\textrm{liberation}\,=\,0.83}$ with an uncertainty between clusters of $\pm\,0.04$. Not all ICL particles are assigned a liberation time because the identification requires a star particle to be a satellite member in one snapshot and a BCG + ICL member in the subsequent snapshot, and so naturally does not include stars of infalling groups that have been liberated from a satellite before it was first seen inside the cutout box. This material would typically be referred to as pre-processed ICL which makes up the intragroup light of the infalling group. Note that pre-processed stars always account for a minority of the ICL in our sample \citep[see][]{Brown_2024}. There may also be contributions from star particles that were formed in a satellite but liberated before the next snapshot, and star particles formed in the BCG that later became part of the ICL, however these channels account for less than 5 per cent \citep{Brown_2024} of the ICL at $z\approx0$. Expanding the satellite definition from $4\,r_\textrm{sat}$ to $8\,r_\textrm{sat}$ will also reduce $f_\textrm{liberation}$. However, we prioritise purity over completeness, so a conservative identification resulting in ${f_\textrm{liberation}<1}$ is not a concern.

\subsection{Defining the orbital properties}\label{sec:orbital_properties}

The orbits of particles can be described in terms of kinetic and potential energies. These quantities vary throughout a particle's orbit, but their sum (the orbital energy) is conserved. A typical DM particle will have a higher orbital energy than a typical star particle simply due to its mass ($M_\textrm{DM}/ M_\textrm{star}\,\approx\,40$), so to allow comparisons we divide the energies by particle mass, and thus use specific kinetic energy ($\varepsilon_\textrm{k}$) and specific potential energy ($\varepsilon_\textrm{p}$). The specific total energy ($\varepsilon$) is defined as

\begin{equation}
   \varepsilon = \varepsilon_\textrm{k} + \varepsilon_\textrm{p} = \frac{v^2}{2} + (\phi - \phi_\textrm{min}),
    \label{eq:specific_energy}
\end{equation}
where $v$ is the particle speed, $\phi$ is the potential at the position of the particle, and $\phi_\textrm{min}$ is the potential at the position of the most bound particle in the cluster. The potential is computed following the \cite{Barnes_Hut_1986} method.\footnote{Our implementation can be found at: \url{https://github.com/garrethmartin/bh_potential}.} The normalisation to $\phi_\textrm{min}$ means the specific orbital energy will always be positive, thus the usual implication that a positive total energy corresponds to an unbound particle does not apply. In a static potential, $\varepsilon$ would be constant throughout the orbit for a given particle. 

We also compute $\varepsilon_\textrm{k}$ and $\varepsilon_\textrm{p}$ for the satellite galaxies, treating them as point masses. The radial position, $r$, and speed, $v$, of each galaxy is calculated using the mean position and velocity of the star particles within $r_\textrm{sat}$ of the galaxy centre. Calculating $\varepsilon_\textrm{k}$ from $v$ is straightforward, but for calculating $\varepsilon_\textrm{p}$, we must approximate the potential of the cluster at the position of the galaxy's centre, since the presence of the galaxy itself will lower the potential at that point. We do this by taking the mean of $\varepsilon_\textrm{p}$ of the DM particles on a thin shell at $r$, excluding particles that are within $8\,r_\textrm{sat}$ of any satellite.

The orbital properties of the intracluster stars and DM are not fully defined by their orbital energy; we also need to examine their angular momenta, which we explore through the orbital anisotropy, given by:

\begin{equation}
   \beta(r) = 1 - \frac{\sigma_\theta(r)^2+\sigma_\phi(r)^2}{2\,\sigma_r(r)^2},
    \label{eq:anisotropy}
\end{equation}
where $\sigma_\theta$, $\sigma_\phi$ and $\sigma_r$ are the velocity dispersions of the sample of particles in spherical coordinates  \citep{Binney_Tremaine_2008}. $\beta(r)=0$ corresponds to a population of isotropic orbits, $\beta(r)=1$ corresponds to a population on purely radial orbits (with $\beta>0$ generally said to be a orbits with a radial bias), whereas $\beta(r)\rightarrow-\infty$ corresponds to circular orbits (with $\beta<0$ describing orbits with a tangential bias).

\begin{figure*}
	\includegraphics[width=\textwidth]{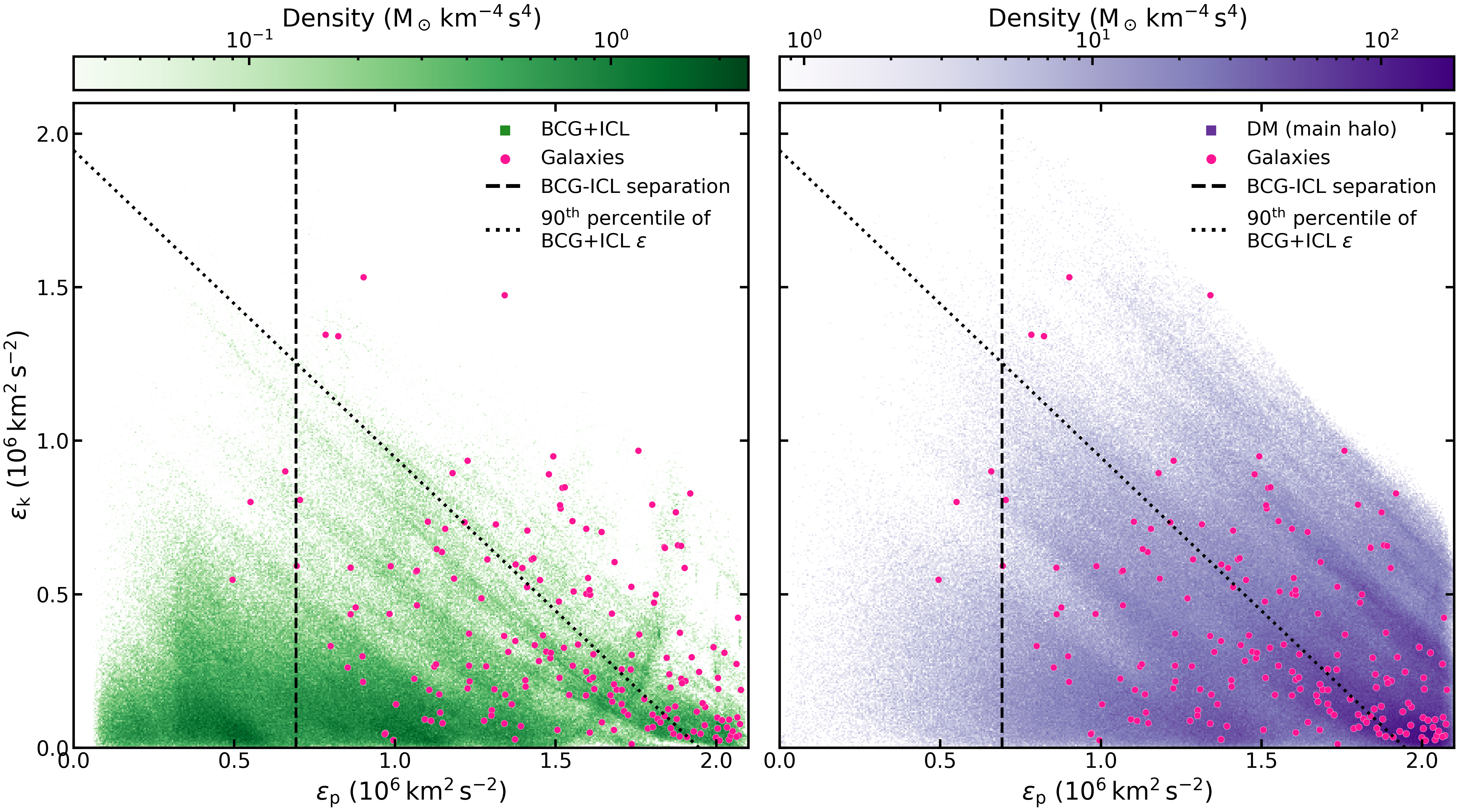}
    \caption{Mass-weighted 2D histogram of specific kinetic energy ($\varepsilon_\textrm{k}$) against specific potential energy ($\varepsilon_\textrm{p}$) for the BCG + ICL (left, green) and main halo DM (right, purple) for an individual cluster, Hrz001. Darker colours represent higher densities on a logarithmic scale. The individual galaxies are plotted as magenta circles, the dashed black line indicates the mean $\varepsilon_\textrm{p}$ at the boundary between the BCG and ICL, and the dotted black line indicates the 90th percentile of the specific (total) energy distribution of the BCG + ICL.}
    \label{fig:phase_space_energies}
\end{figure*}

\section{Results}\label{sec:results}

\subsection{Energetics}\label{sec:energetics}

\begin{figure}
	\includegraphics[width=\columnwidth]{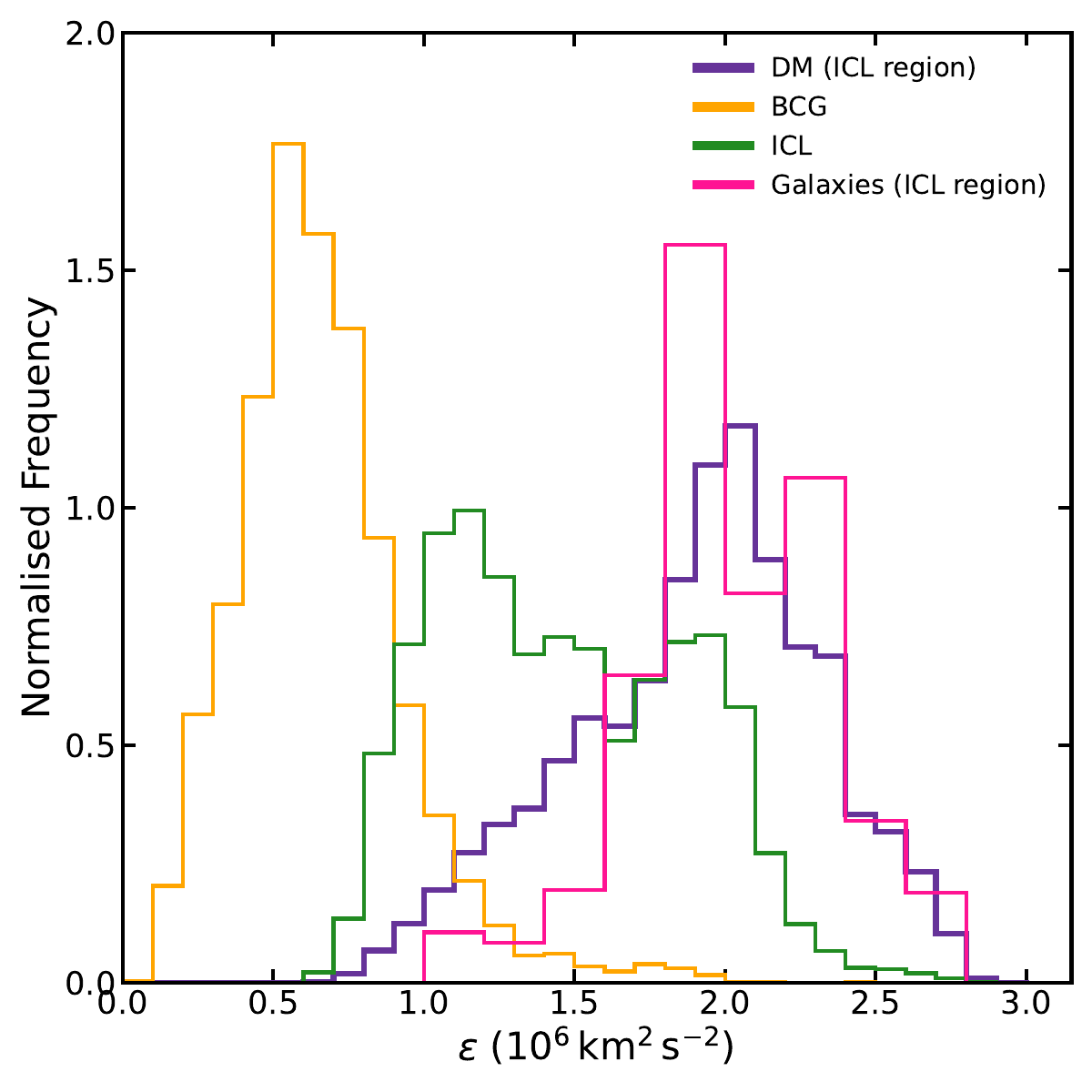}
    \caption{Specific energy ($\varepsilon$) distribution for DM in the ICL region (purple), BCG (orange), ICL (green) and galaxies in the ICL region (magenta) for an individual cluster, Hrz001.}
    \label{fig:se_distribution}
\end{figure}

We begin by examining the orbital energetics of a single cluster, Hrz001, at $z\approx0$. In Fig.\,\ref{fig:phase_space_energies} we show the energetics of the stellar particles of the BCG and ICL components (left panel) and the main halo DM (right panel), with the galaxies overlaid in magenta. We define the phase space as the specific potential energy versus the specific kinetic energy. The galaxies preferentially sample almost the same regions of phase space as the DM. Both populations preferentially occupy the higher potential energy and lower kinetic energy regions of the phase space as any non-circular orbit will spend more time near its apocentre, where potential energy is at maximum, than its pericentre.

Conversely, the distribution of the BCG + ICL stellar particles is fundamentally different to the DM in this phase space: BCG + ICL particles have, on average, even higher potential energies relative to their kinetic energy than the DM particles, indicating a larger radial bias. The majority of the BCG + ICL particles also occupy overall lower specific energies than the DM particles. To demonstrate this mismatch, the $90^\textrm{th}$ percentile of the BCG + ICL specific orbital energy distribution is added to both panels as a diagonal, black dotted line. The significant fraction of DM particles above this line (48 per cent) clearly demonstrates that the ICL poorly samples the full phase space occupied by the DM energy distribution. We note that this mismatch in energetics is not just dominated by the stars associated with the BCG: the vertical, dashed line indicates the approximate boundary between the BCG and ICL, which we define to be at 100\,kpc, and there is a large difference in the phase space distribution between the ICL and the DM to the right of that line (in the ICL region).

Clear structure is visible in both the BCG + ICL and DM distributions, much of which is aligned along isoenergy surfaces (diagonal lines in Fig.\,\ref{fig:phase_space_energies}). Some of these will be identifiable coherent structures in 6-D phase space such as stellar streams \citep{Rudick_2009}, whereas others will have undergone significant phase mixing. Further investigation of these structures is however beyond the scope of this paper.

We plot the specific total energy distribution of the BCG stars, ICL stars, DM and galaxies from cluster Hrz001 in Fig.\,\ref{fig:se_distribution} to show that the difference in the energetics of the DM and ICL stars is not a simple translation in energy states. We focus on the energetics of the star particles belonging to the ICL and compare them to the DM and galaxies that are located in the ICL region (i.e. at radii greater than 100\,kpc from the cluster centre, as defined in Section\,\ref{sec:defining_bcg_icl}).

The specific energies of the galaxy population are in reasonable agreement with the DM, except for a deficiency of galaxies at the lowest specific energies ($<1.5\times10^{6}\,\textrm{km}^2\,\textrm{s}^{-2}$). The sharp decline in the distributions of both DM and galaxy components beyond $\sim2.1\times10^{6}\,\textrm{km}^2\,\textrm{s}^{-2}$ results from truncating the cluster boundary at $R_{178}$. However, the ICL stars display a very different distribution to the DM, with most stars having low specific energies ($<1.5\times10^{6}\,\textrm{km}^2\,\textrm{s}^{-2}$). The sharp decline at $\sim0.7\times10^{6}\,\textrm{km}^2\,\textrm{s}^{-2}$ is due to the separation between the BCG and ICL that occurs at 100\,kpc, which results in the ICL stars having a minimum specific total energy of the $\varepsilon_\textrm{p}$ at 100\,kpc. We emphasise that the precise shapes of these distributions therefore depend on these cuts at 100\,kpc and $R_{178}$, and also varies between clusters. For example, the apparent bimodality seen in the distribution of the ICL is not a consistent feature. However, the broad result that the galaxy distribution is similar to the DM distribution, and the ICL distribution mostly populates lower energies, holds.

\begin{figure}
	\includegraphics[width=\columnwidth]{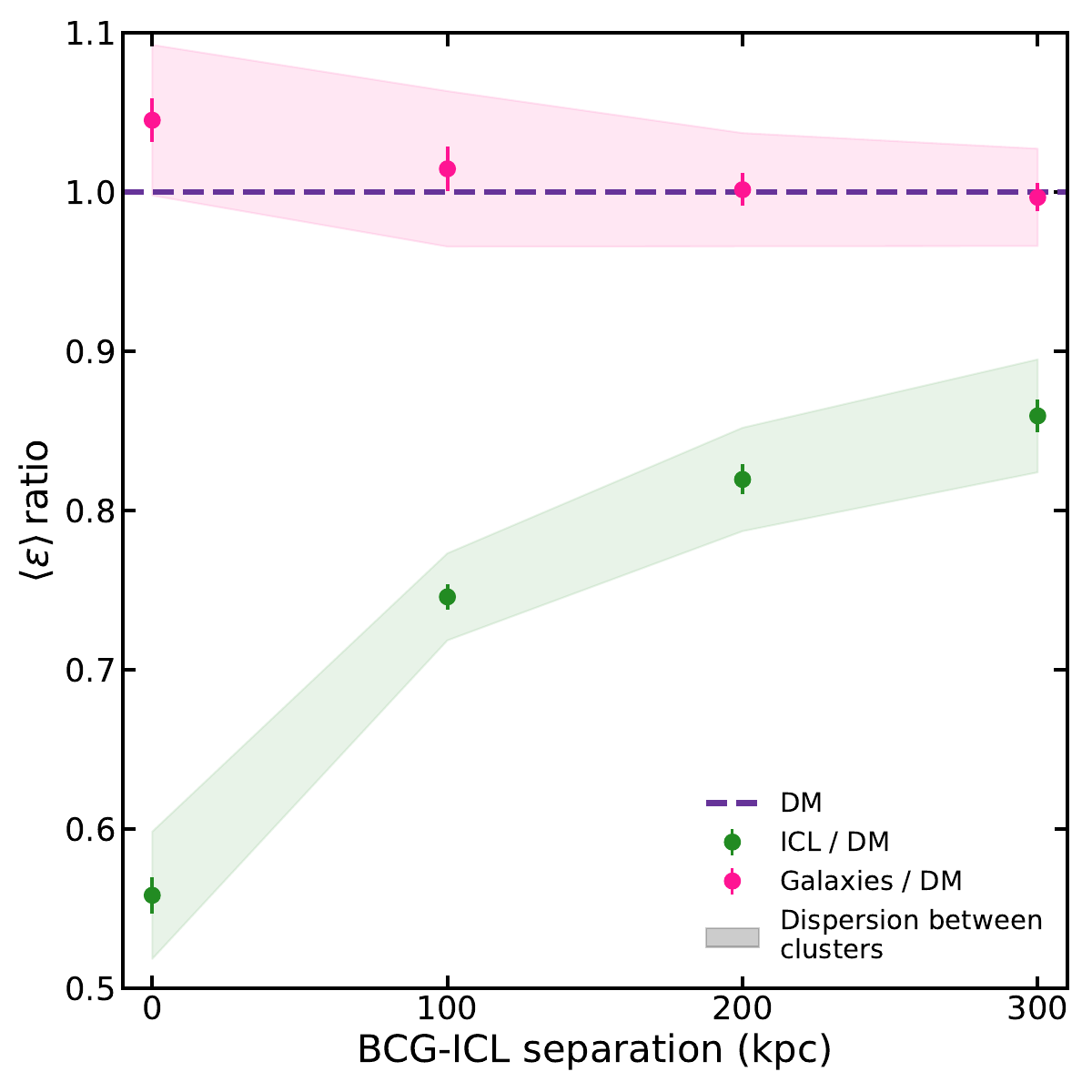}
    \caption{Specific energy ratio of galaxies ($\langle \varepsilon_\textrm{galaxies/DM} \rangle$, magenta) and ICL ($\langle \varepsilon_\textrm{ICL/DM} \rangle$, green) at different BCG--ICL separations, such that only the populations beyond this separation are considered for all components (galaxies, ICL and DM). The points represent the means, the error bars are the $1\,\sigma$ standard error on the mean, the shaded regions are the $1\,\sigma$ dispersion between clusters, and the purple dashed line indicates a specific energy ratio of 1. The extreme BCG--ICL separation of 300\,kpc still resulting in ${\langle \varepsilon_\textrm{ICL/DM} \rangle < 1}$ demonstrates that the BCG is not the primary driver of this difference in average specific energies.}
    \label{fig:ser_separations}
\end{figure}

We next explore the orbital energies in all 12 of the simulated clusters at $z\approx0$. The total energy for each cluster varies due to the different cluster masses, therefore, to normalise the orbital information of all the clusters we first compute the specific energy ratio $\langle \varepsilon_\textrm{component/DM} \rangle$, defined as the average specific energy $\langle \varepsilon \rangle$ for each component (BCG + ICL and galaxies) divided by that of the DM for each cluster. Finally we calculate the average of these ratios for all 12 clusters. In this calculation a mass-weighted average is taken for stellar and DM components, while galaxies are treated as point tracers, with no mass-weighting. We calculate the specific energy ratios for all cluster components within $R_{178}$, but also calculate them while excluding the inner 100\,kpc, 200\,kpc and 300\,kpc to explore the impact of including the star particles that are part of the BCG. These specific energy ratios are shown in Fig.\,\ref{fig:ser_separations}. 

Consistent with Figs. \ref{fig:phase_space_energies} and \ref{fig:se_distribution}, the energetics of the galaxies and the DM halo are similar, with $\langle \varepsilon_\textrm{galaxies/DM} \rangle = 1.05\,\pm\,0.01$ (as seen in Fig.\,\ref{fig:ser_separations}). The quoted uncertainties of specific energy ratios are the $1\,\sigma$ standard error on the mean, but we also note that the dispersion between clusters (the shaded regions in Fig.\,\ref{fig:ser_separations}) is consistently $\leq\,0.05$. The slightly higher energies of the galaxies compared to the DM halo can be explained by galaxies with low orbital energy preferentially merging with the BCG and thus being effectively removed from the galaxy sample \citep[galactic cannibalism,][]{Ostriker_1975}. Indeed, if only the DM and galaxies beyond 100\,kpc are considered, then this ratio drops to $\langle \varepsilon_\textrm{galaxies/DM} \rangle = 1.01\,\pm\,0.01$, and remains consistent with 1 when this cut is increased to 200\,kpc and 300\,kpc. This demonstrates that the satellite galaxies sample the DM halo phase space well and are thus an effective, mostly-unbiased tracer of the underlying DM.

If the intracluster stars were unbiased tracer particles of the DM halo, we would expect $\langle \varepsilon_\textrm{ICL/DM} \rangle$ to be approximately unity. However, taking the ICL region to be that beyond 100\,kpc of the BCG, we find that $\langle \varepsilon_\textrm{ICL/DM} \rangle = 0.75\pm0.01$, i.e. the mean specific energy of the ICL is only $\approx\,75\,$per cent of the DM.

The stars associated with the BCG are expected to have lower energies than those associated with the ICL due to their location near the bottom of the potential well. Therefore, when we include all the BCG stars with the ICL component, we measure $\langle \varepsilon_\textrm{BCG + ICL/DM} \rangle = 0.56\pm0.01$. As previously mentioned, \cite{Proctor_2023} identified the BCG--ICL transition region as $\sim100\,\textrm{kpc}$ for $10^{14}\,\textrm{M}_{\odot}$ haloes, so excluding the central 100\,kpc region may still leave a contribution of star particles that belong to the BCG outskirts in this definition of the ICL. In Fig.\,\ref{fig:ser_separations} we show that ${\langle \varepsilon_\textrm{ICL/DM} \rangle}$ increases as we progressively exclude more of the core of the cluster. Importantly, however, even in the extreme case of excluding out to a radius of 300\,kpc (at which point any BCG contribution can safely be assumed negligible), we find that ${\langle \varepsilon_\textrm{ICL/DM} \rangle}$ still only reaches $0.86\,\pm\,0.01$, meaning the contamination from the BCG is not the primary driver of our results. Overall, these show that although the separation between the BCG and ICL will affect the precise ${\langle \varepsilon_\textrm{ICL/DM} \rangle}$, the general result that ${\langle \varepsilon_\textrm{ICL/DM} \rangle < 1}$ holds.

We also test the robustness of these results to different definitions of satellite galaxies. The factor of 4 in the spherical cut of radius ${4\,r_\textrm{sat}}$ around the centre of each galaxy is relatively arbitrary, so we test factors of 6, 8 and 10. The effect on ${\langle \varepsilon_\textrm{ICL/DM} \rangle}$ is a negligible increase, only increasing from $0.75\pm0.01$ for a factor of 4 to $0.77\pm0.01$ for a factor of 10. However, increasing this factor results in significant decreases of BCG + ICL mass as more stars are attributed to galaxies. Thus, we choose to continue using a factor of 4.

\subsection{Anisotropy}\label{sec:anisotropy}

\begin{figure}
	\includegraphics[width=\columnwidth]{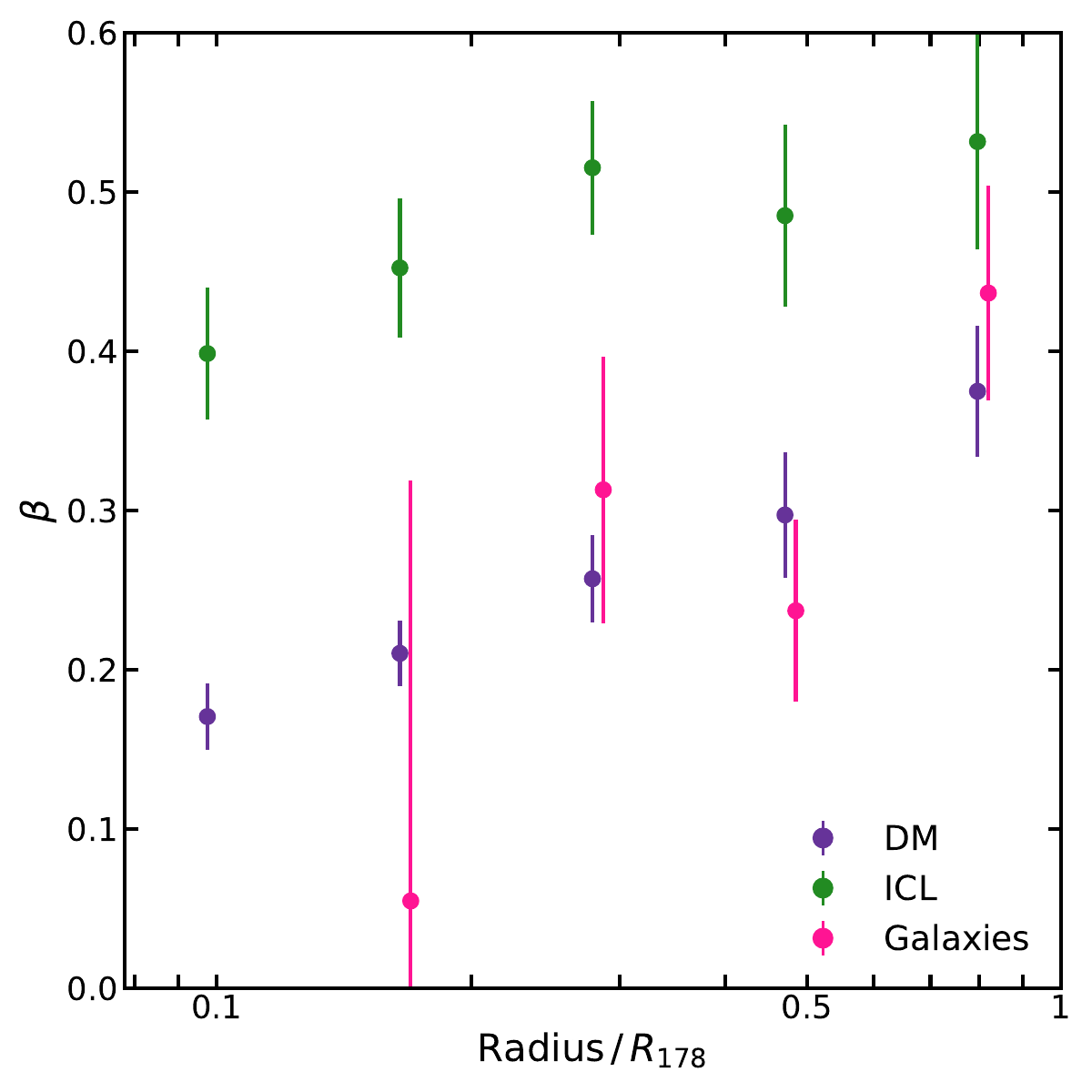}
    \caption{Anisotropy $\beta$ against radius for dark matter (purple), ICL (green) and galaxies (magenta). The points represent the mean between clusters, and the errors bars are the $1\,\sigma$ standard error on the mean. The midpoints of the radial bins for the galaxies are plotted slightly offset for visual clarity, and truncated below $\approx0.15\,R_\textrm{178}$ due to low counts.}
    \label{fig:anisotropy}
\end{figure}

Now that we have established differences between the orbital energetics of the DM, ICL stars and galaxies, we turn our attention to the orbital anisotropy. Fig.\,\ref{fig:anisotropy} shows the anisotropy profiles of the DM, ICL stars and galaxies averaged between clusters.

We see that the DM anisotropy increases (becomes more radial) with cluster-centric radius, increasing from $\beta\approx0.15$ to $\beta\approx0.35$, in agreement with previous studies \citep{Willman_2004, Wojtak_2005, Ascasibar_2008, Lemze_2012, He_2024}. The anisotropy profile of the galaxies (magenta points) is truncated below $\approx0.15\,R_\textrm{178}$ due to low counts, and has a larger dispersion between clusters, such that a positive trend with radius is less certain. However, the average galaxy anisotropy profile from all 12 clusters is similar to the DM profile, in agreement with figure 10 from \cite{Willman_2004}. Our average galaxy anisotropy profile is in qualitative agreement with observational studies \citep{Biviano_2013, Mamon_2019} that find $\beta\approx0$ in the cluster core to $\beta\approx0.5$ at $R_{200}$.

Other studies investigate the anisotropy profile of the subhaloes \citep{Diemand_2004, He_2024} and find the subhalo anisotropy is typically lower than in our work. For example, \cite{He_2024} find the anisotropy of subhaloes increases from $\beta\approx0$ to $\beta\approx0.2$ in the halo mass range $\textrm{log}_{10}\,\textrm{(}M_\textrm{200}/\textrm{M}_\odot\textrm{)} = 13.8\,-\,14.3$. However, it is important to note that the anisotropy profiles of the galaxies and subhaloes are not directly comparable since orphan galaxies exist in some simulations.

An increase in anisotropy towards more radial orbits can also be seen for the ICL stars, increasing from $\beta\approx0.4$ to $\beta\approx0.55$. This is in good agreement with \cite{He_2024} (see their figure 6), who also find the same trend and similar values for accreted halo stars in the TNG300 simulation. \cite{Willman_2004} also find that $\beta$ increases with radius for unbound stars, albeit encompassing a much larger range, increasing from $\beta\approx-0.2$ to $\beta\approx0.8$. Importantly, we find that at all radii, the anisotropy of the ICL stars is more radially biased than that of the DM and galaxies.

\section{Discussion}\label{sec:discussion}

Our work demonstrates that the orbital energies and anisotropy of the ICL stars inherently differ from both the galaxies and the DM. This is despite the dynamics of each of these components being governed by the same potential, and the ICL stars primarily originating from galaxies. We discuss potential reasons why the orbital properties of the ICL stars are different to the satellite galaxies and the DM, and the observable implications of these differences.

\subsection{Why do the orbits differ?}\label{sec:reasons}
\begin{figure}
	\includegraphics[width=\columnwidth]{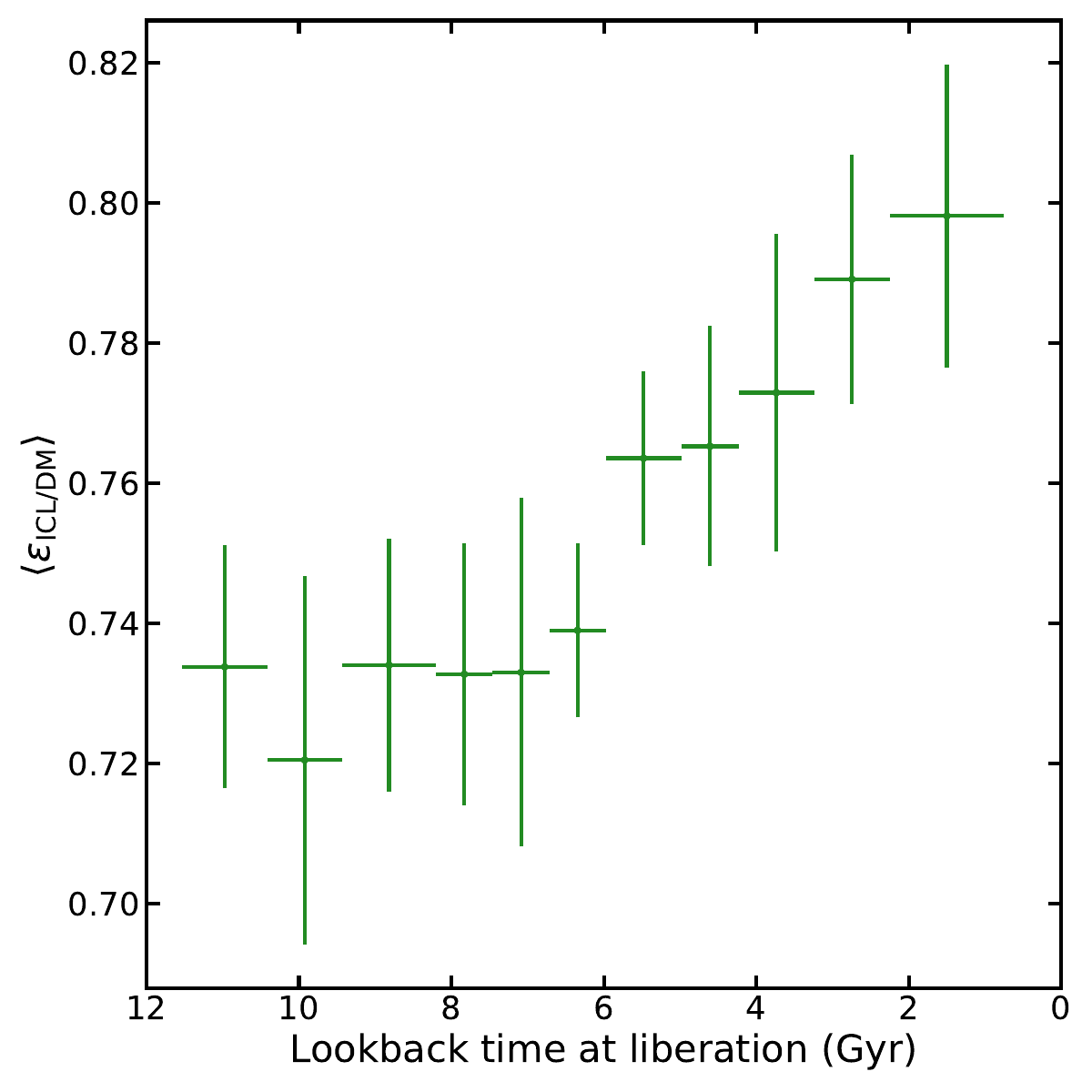}
    \caption{Specific energy ratio $\langle \varepsilon_\textrm{ICL/DM} \rangle$ split up by liberation time. This is the $\langle \varepsilon_\textrm{ICL} \rangle$ at $z\approx0$ of intracluster stars with a given liberation time, divided by $\langle \varepsilon_\textrm{DM} \rangle$ at $z\approx0$. The points represent the mean between clusters, the vertical error bars the $1\,\sigma$ standard error on the mean, and the horizontal error bars the width of the individual age bins. This plot demonstrates that even the most recently liberated stars have $\langle \varepsilon_\textrm{ICL/DM} \rangle < 1$.}
    \label{fig:ser_liberation}
\end{figure}

\subsubsection{Evolving cluster potential}

We first consider the possibility that the energies of the intracluster stars could be lower than the DM because the intracluster stars' energy reflects the evolving properties of the cluster potential at the time of stripping. Galaxies that entered the cluster at earlier times, when it was less massive, would travel on lower energy orbits compared to galaxies infalling at later times, and thus the orbits of the stars liberated through tidal stripping would also have these lower energies. If these orbital energies at liberation time were approximately maintained until ${z=0}$, one might expect that $\langle \varepsilon_\textrm{ICL/DM} \rangle$ split up by liberation time would steadily increase towards ${z=0}$, with the most recently liberated material matching the energetics of the ${z=0}$ DM halo. However, in Fig.\,\ref{fig:ser_liberation} we show there is no clear dependence of $\langle \varepsilon_\textrm{ICL/DM} \rangle$ on liberation time until a lookback time of $\sim6\,\textrm{Gyr}$, after which there is a slight increase in $\langle \varepsilon_\textrm{ICL/DM} \rangle$, such that more recently liberated stars tend to have higher energies. However, this trend is not consistent between different definitions of satellite galaxies and BGC-ICL separations, so we do not explore this further. More importantly, even the most recently liberated stars have $\langle \varepsilon_\textrm{ICL/DM} \rangle = 0.80\,\pm\,0.02$, and thus do not reach the average energies of the DM. This mechanism, therefore, cannot fully explain the lower energetics of the ICL stars compared with the DM.

\subsubsection{Differential stellar and DM stripping}

An alternative possibility is that a majority of intracluster stars are stripped from galaxies on orbits with lower energies and/or higher angular momenta than the DM. Many studies have found that the ICL is not built up by uniform contributions from all galaxies \cite[e.g.][]{Contini_2014, Contini_2019, Chun_2023, Chun_2024, Ahvazi_2024b, Brown_2024} and this also applies to their orbits. As the DM halo of an orbiting satellite is, on average, more weakly bound than its stars, significant amounts of DM can be tidally stripped on orbits that result in very little stellar stripping \citep[e.g.][]{Villalobos_2012, Smith_2016b, Joshi_2019, Haggar_2021, Martin_2024}. This would therefore bias the orbital energies and anisotropy of the intracluster stars to those of galaxies on lower energy and more radial orbits, and thus explain why the intracluster stars have lower specific energies (Figs.\,\ref{fig:phase_space_energies} and \ref{fig:se_distribution}) and appear more radially-biased (Fig.\,\ref{fig:anisotropy}) compared with both the DM and the galaxies.

Another possibility is that loss of orbital energy over time due to dynamical friction \citep{Chandrasekhar1943} will result in higher levels of stellar stripping and therefore bias the orbital energies of the intracluster stars to reflect those of more massive galaxies that have fallen into lower orbits. At the same time, galaxies experiencing high levels of dynamical friction will eventually merge with the BCG \citep{Tormen1997}, biasing the satellite population to modestly higher orbital energies \citep{Ostriker_1975, Dolag_2010}. However, this mechanism alone cannot account for the more radial orbits observed in the intracluster stars, as dynamical friction typically results in either no change or in the circularization of satellite orbits, dependent on the density profile of the host cluster \citep{van_den_Bosch_1999,Arena_2007}.

To summarise, our result that the intracluster stars have lower orbital energies and a more radially-biased anisotropy compared to the galaxies and DM particles could be explained by biases in orbits and galaxies that progenerate the bulk of the ICL stars. ${\langle \varepsilon_\textrm{ICL/DM}\rangle}$ may also depend on how the ICL formed in a particular cluster. For example, if a large fraction of ICL was formed from pre-processed material, which is loosely-bound to the infalling subhalo, then the orbital energy difference in such a cluster may be smaller compared to clusters where the ICL formed primarily by mergers with the BCG. ${\langle \varepsilon_\textrm{ICL/DM}\rangle}$ and the radially-biased $\beta$ likely depend on the specific ICL formation process of the cluster, and therefore there may be a scatter in these parameters within the cluster population.

\subsection{Observable implications}
\subsubsection{Density Profile}\label{sec:density_profile}

\begin{figure}
	\includegraphics[width=\columnwidth]{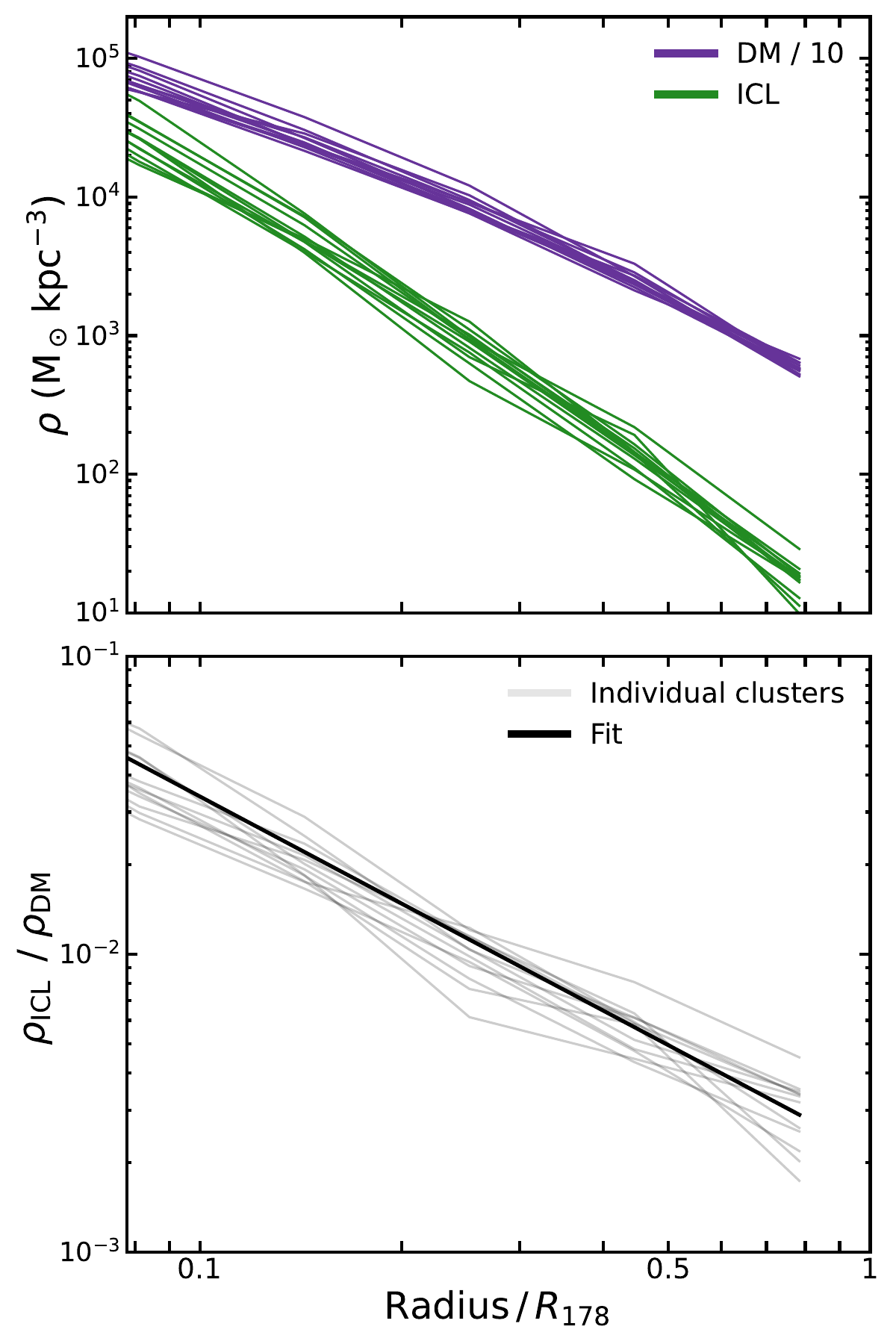}
    \caption{Top panel: Density profiles of DM (purple) and ICL (green). The DM density profile is divided by an arbitrary factor of 10 to make visual comparison of the DM and ICL profile shapes clearer. Bottom panel: Density ratio $\rho_\textrm{ICL}$ / $\rho_\textrm{DM}$ as function of radius. Grey lines are individual clusters and the black line is the power law fit $y=a\cdot x^b$ with $a=(2.16\pm0.12)\times10^{-3}$ and $b=-1.19\pm0.04$.}
    \label{fig:density_ratio}
\end{figure}

The nature of DM can be probed by analysing the radial mass distribution of the halo \citep[e.g.][]{Spergel_2000}, and previous studies \citep[e.g.][]{Alonso_Asensio_2020, Contreras_Santos_2024} have suggested that the ICL radial profile could be used to infer the DM radial profile. However, a notable consequence of the differing energetics and orbital anisotropy is that the density profile of the ICL should differ from the DM profile. The Jeans equation tells us that there are two competing effects: the greater anisotropy of the intracluster stars means the ICL density profile should be flatter than the DM, but the lower orbital energetics of the intracluster stars means the ICL profile should be more concentrated. In Fig.\,\ref{fig:density_ratio} we show the individual density profiles of both the DM and ICL for each cluster (top panel) and their ratio (bottom panel). We see that the ratio of the density profiles between the ICL and DM can be fit by a power law with a negative exponent, such that the density of the ICL decreases faster than the DM. This demonstrates that the lower energetics of the ICL is the dominant factor over the anisotropy in determining the radial profile. The gradient of the density ratio is similar for all clusters, but the normalisation differs. We test if this normalisation is related to ${\langle \varepsilon_\textrm{ICL/DM}\rangle}$ and find no clear correlation.

A comparison of the ICL profile and the DM profile, as measured by gravitational lensing, has been performed for SMACS0723 \citep{Diego2023} and MACS0416 \citep{Diego_2024}. They find that the ICL profile is steeper than the DM profile, in qualitative agreement with our findings. Similarly, \cite{Chen_2022} find the surface mass density profile of the BCG + ICL is more centrally concentrated than the DM.

It should be emphasised that the difference in the shapes (density profiles) of the ICL and DM does \textit{not} imply that their morphologies should be different. Since the dynamics of both components are governed by the same cluster potential, it is perhaps unsurprising that both observations \citep{Montes_Trujillo_2019} and simulations \citep{Alonso_Asensio_2020} agree that the morphologies match well in the inner region, but crucially this does not guarantee that the two components sample this potential in a similar way.

Our work confirms the findings of \citet{Alonso_Asensio_2020} and \citet{Contreras_Santos_2024} who demonstrated that the ratio of the density profiles of the intracluster stars and DM approximately follows a power law relationship. However, the exact mechanisms driving this relationship remain unclear. It is uncertain whether this relationship holds universally across different cluster properties such as dynamical states, mass and redshift. Furthermore, the observed differences between the density profiles of DM and ICL may be sensitive to the dominant ICL formation process as well as the properties of the galaxy populations that progenerate the ICL, whose size and stellar mass distribution are themselves sensitive to the implemented galaxy evolution models. Finally, the presence of non-phase-mixed material may cause temporal variations in the density profiles. We therefore caution against using this relation to estimate the DM profile from the observed ICL profile, until the reasons for the differing energetics and orbital anisotropies are better understood.

\subsubsection{Kinematics}\label{sec:kinematics}

\begin{figure*}
	\includegraphics[width=\textwidth]{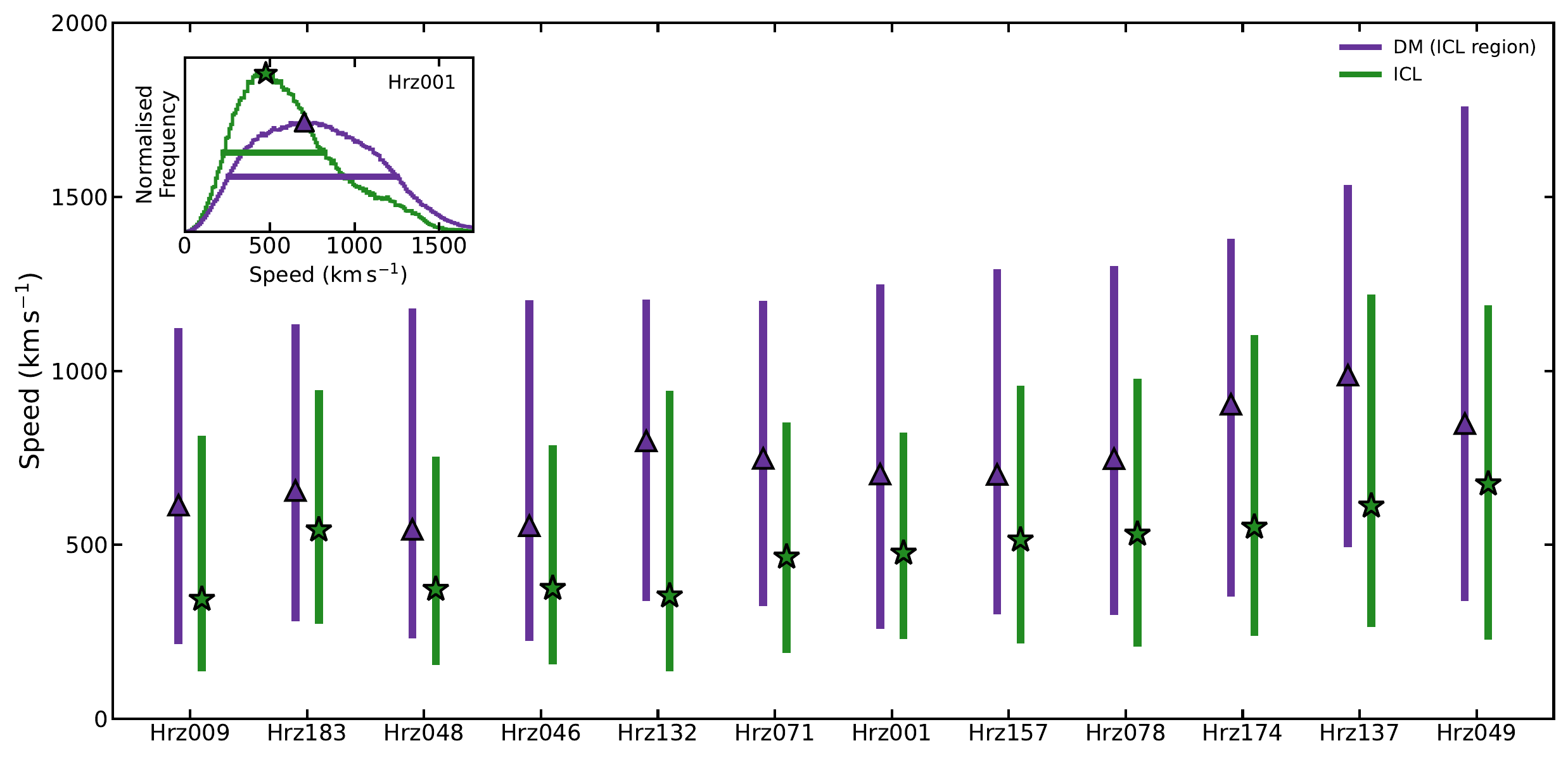}
    \caption{Velocity distributions of the DM in the ICL region (purple) and ICL (green) for each cluster, ordered by ${z=0}$ mass. The peaks of the velocity distributions are represented by triangles for the DM and stars for the ICL, and the error bars represent the full width at half maximum of the velocity distributions. The inset indicates these quantities for Hrz001. The difference in the peaks of the velocity distributions for each cluster are given in Table\,\ref{tab:cluster_properties}.}
    \label{fig:vds}
\end{figure*}

Another potentially observable implication of the differing energetics of the ICL and DM is that their velocity distributions will also differ. The key information about the velocity distributions for each cluster in order of increasing $M_{178}$ are summarised in Fig.\,\ref{fig:vds}. For each cluster, the arrays of DM and ICL particle speeds are bootstrapped 100 times and splines fit to each. The peak of each distribution (represented by triangles for DM and stars for ICL) is then computed from the mean of the peaks of each spline, and the full width at half maximum (FWHM, represented by solid bars) from where the median curve through these splines equals half the peak height. The large number of particles in both components ensures the error on the mean speed of their spline peaks are negligible. The top-left inset illustrates this for one of the clusters.

For every cluster, the peak speed is higher for the DM than the ICL stars. This is in agreement with \cite{Contreras_Santos_2024}, who use a similar aperture-based method of BCG--ICL separation with a 50\,kpc radius sphere, and find the velocity dispersion ratio between the ICL and DM to be $\approx 0.75$ within $R_{500}$. Notably, our work also agrees with \cite{Dolag_2010}, who employ a kinematic separation of the BCG and ICL. This method naturally selects the higher speed stars to be associated with the ICL instead of the BCG, and yet the velocity distribution of the ICL still peaks at a lower speed than the DM.

Observationally, measuring the kinematics of the ICL is extremely difficult, due to both its faint nature and uncertainty in the transition radius. Integrated-light absorption spectra has been measured for many BCGs \citep[e.g.][]{Carter_1981, Carter_1985, Kelson_2002, Bender_2015, Boardman_2017, Barbosa_2018, Loubser_2022}, but typically the kinematics are limited to within $3-4$\,half-light radii. Some studies \citep[e.g.][]{Bender_2015, Edwards_2020} have found that the stellar velocity dispersion reaches that of the cluster in the BCG outskirts, which may disagree with our result that the velocity dispersion of the ICL should never reach that of the DM. However, the difference in velocity dispersions between the two components may not be pronounced, and therefore be within uncertainties. For example, \cite{Dolag_2010} find a $\sim500\,\textrm{km}\,\textrm{s}^{-1}$ difference in the peak speeds of the DM and the diffuse stellar component, but this translates to a difference in velocity dispersion of only $\sim20\,\textrm{km}\,\textrm{s}^{-1}$ (see their figure 5).

To probe the ICL at larger distances, most studies attempt to identify and measure the kinematics of bright tracers, usually intracluster planetary nebulae \citep[PNe,][]{Arnaboldi_1998, Longobardi_2013, Hartke_2017, Pulsoni_2018} and/or globular clusters \citep[GCs,][]{Schuberth_2010, Strader_2011, Pota_2018}. There have only been a handful of successful observations to date, which include the Virgo, Fornax, Hydra I and Coma clusters -- see \cite{Arnaboldi_2022} and references therein for full detail on these. However, different ICL tracers may have different energetics and anisotropy than the global intracluster stars. Simulations \citep{Reina-Campos_2022} and observations \citep{Alabi_2016, Alabi_2017, Hudson_2018} show that intracluster GCs are good tracers of dark matter haloes, and so may have energetics and anisotropy more similar to the DM rather than the ICL. On the other hand, \citet{Ramos-Almendares2020} simulate intracluster GCs by associating GCs with DM particles selected to match the radial distribution of GCs in observed galaxies. They show that the GC kinematics are 25 per cent lower than the DM, which aligns with our results for the ICL. 

\section{Conclusions}\label{sec:conclusions}

In this study we have investigated the energetics of the ICL and DM at $z\approx0$, using 12 clusters with masses $1.18 - 3.71 \times 10^{14}\,\textrm{M}_\odot$ from the {\sc Horizon-AGN} cosmological hydrodynamic simulation. We summarise our results below:

\begin{enumerate}
    \item We calculate the mean specific energies ${\langle \varepsilon \rangle}$ of the DM halo, ICL, and satellite galaxies. We find that the ratio between that of the ICL and DM is ${\langle \varepsilon_\textrm{ICL/DM} \rangle = 0.75\pm0.01}$, such that the average orbital energy of the intracluster stars is $\approx75$\,per cent of that the DM.
    On the other hand, galaxies (treated as point masses) have marginally higher energies than the DM with ${\langle \varepsilon_\textrm{galaxies/DM} \rangle = 1.05\,\pm\,0.01}$. Whilst the specific energy distributions of the DM and galaxies appear to match well (Fig.\,\ref{fig:se_distribution}), the ICL has a very different distribution, concentrated towards lower energies and poorly sampling the higher energies at which much of the DM and galaxies reside (Fig.\,\ref{fig:phase_space_energies}).
    \item These results are tested against different BCG--ICL separations (Fig.\,\ref{fig:ser_separations}). The energetics of the galaxy population becomes consistent with the DM halo when the innermost 100\,kpc of both components are excluded (${\langle \varepsilon_\textrm{galaxies/DM} \rangle = 1.01 \pm0.01}$), showing that there is a deficit in the low-energy galaxy population in the cluster core, but the galaxies are otherwise a good tracer of DM. As the BCG--ICL separation is increased to an extreme of 300\,kpc, where the BCG contribution can safely be assumed to be negligible, ${\langle \varepsilon_\textrm{ICL/DM} \rangle}$ still only reaches $0.86\,\pm\,0.01$. This demonstrates that the lower energetics of the ICL is not simply due to BCG contribution, and thus this qualitative result is robust to the precise definition of the ICL. We also test the impact of expanding the extent of satellite galaxies, which assigns additional stars to the satellites galaxies instead of to the ICL, and find only small quantitative differences.
    \item The angular momentum of the cluster components is quantified by the orbital anisotropy. The ICL is found to be more radially biased (larger $\beta$) than the DM at all cluster-centric radii (Fig.\,\ref{fig:anisotropy}). This, in combination with the differing energetics, highlights that the orbital configurations of the ICL and DM are inherently different.
    \item These differences in the energetics and anisotropy of the ICL and DM manifest themselves in their velocity distributions and density profiles. We find that the ICL density profile is more centrally-concentrated than that of the DM (Fig.\,\ref{fig:density_ratio}), such that the ratio of their densities as a function of radius has a negative gradient. We also find that the peak speed is lower and FWHM is smaller for the ICL velocity distribution than that of the DM for all clusters (Fig.\,\ref{fig:vds}). Both of these results qualitatively agree with other simulation work.
\end{enumerate}

While ICL represents a promising tool for exploring the properties of DM, we have shown that it is not an unbiased proxy for the underlying DM distribution. ICL stars have, on average, lower orbital energies and more radial orbits than DM, meaning their density profiles and velocity distributions differ. The relationship between ICL and DM likely also depends on other factors, such as cluster mass, dynamical state, and the interplay of different mechanisms driving ICL production. Given the limitations of traditional methods like weak lensing, incorporating ICL as a complementary tracer could significantly improve our ability to map DM, particularly with upcoming surveys such as LSST and {\it Euclid}, which will provide deep imaging for large numbers of clusters. A comprehensive understanding of the processes that generate ICL -- and the origins of its discrepancies with DM -- will therefore be essential for calibrating its use effectively.

\section*{Acknowledgements}

We thank the anonymous referee for their constructive comments which have improved this work. We also thank members of the NottICL group for helpful discussions and comments, in particular Yannick Bahé, Jesse B. Golden-Marx, Tutku Kolcu and Harley J. Brown.

J.~B  and N.~A.~H gratefully acknowledge support from the Leverhulme Trust through a Research Leadership Award. G.~M, F.~R.~P, and N.~A.~H acknowledge support from the UK STFC under grant ST/X000982/1. S.~B acknowledges funding support from the Australian Research Council through a Discovery Project DP190101943. This research was supported by the International Space Science Institute (ISSI) in Bern, through ISSI International Team project \#23-577.

This work was granted access to the HPC resources of CINES under allocations 2013047012, 2014047012, and 2015047012 made by GENCI. This work has made use of the Infinity cluster, hosted by the Institut d’Astrophysique de Paris. We warmly thank S. Rouberol for running it smoothly.

\section*{Data Availability}

The raw data products of the {\sc Horizon-AGN} simulation are available upon reasonable request through the collaboration’s website: \url{https:// www.horizon-simulation.org/}. The data products generated from this work are available upon reasonable request from the corresponding author.


\bibliographystyle{mnras}
\bibliography{paper} 

\begin{thebibliography}{}
\makeatletter
\relax
\def\mn@urlcharsother{\let\do\@makeother \do\$\do\&\do\#\do\^\do\_\do\%\do\~}
\def\mn@doi{\begingroup\mn@urlcharsother \@ifnextchar [ {\mn@doi@} {\mn@doi@[]}}
\def\mn@doi@[#1]#2{\def\@tempa{#1}\ifx\@tempa\@empty \href {http://dx.doi.org/#2} {doi:#2}\else \href {http://dx.doi.org/#2} {#1}\fi \endgroup}
\def\mn@eprint#1#2{\mn@eprint@#1:#2::\@nil}
\def\mn@eprint@arXiv#1{\href {http://arxiv.org/abs/#1} {{\tt arXiv:#1}}}
\def\mn@eprint@dblp#1{\href {http://dblp.uni-trier.de/rec/bibtex/#1.xml} {dblp:#1}}
\def\mn@eprint@#1:#2:#3:#4\@nil{\def\@tempa {#1}\def\@tempb {#2}\def\@tempc {#3}\ifx \@tempc \@empty \let \@tempc \@tempb \let \@tempb \@tempa \fi \ifx \@tempb \@empty \def\@tempb {arXiv}\fi \@ifundefined {mn@eprint@\@tempb}{\@tempb:\@tempc}{\expandafter \expandafter \csname mn@eprint@\@tempb\endcsname \expandafter{\@tempc}}}

\bibitem[\protect\citeauthoryear{{Ahvazi}, {Sales}, {Navarro}, {Benson}, {Boselli}  \& {D'Souza}}{{Ahvazi} et~al.}{2024a}]{Ahvazi_2024a}
{Ahvazi} N.,  {Sales} L.~V.,  {Navarro} J.~F.,  {Benson} A.,  {Boselli} A.,   {D'Souza} R.,  2024a, \mn@doi [Open J. Astrophys.] {10.33232/001c.127132}, 7, 111

\bibitem[\protect\citeauthoryear{Ahvazi, Sales, Doppel, Benson, D’Souza  \& Rodriguez-Gomez}{Ahvazi et~al.}{2024b}]{Ahvazi_2024b}
Ahvazi N.,  Sales L.~V.,  Doppel J.~E.,  Benson A.,  D’Souza R.,   Rodriguez-Gomez V.,  2024b, \mn@doi [\mnras] {10.1093/mnras/stae848}, 529, 4666

\bibitem[\protect\citeauthoryear{Aihara et~al.,}{Aihara et~al.}{2018}]{Aihara_2018}
Aihara H.,  et~al., 2018, \mn@doi [\pasj] {10.1093/pasj/psx066}, 70, S4

\bibitem[\protect\citeauthoryear{Aihara et~al.,}{Aihara et~al.}{2022}]{Aihara_2022}
Aihara H.,  et~al., 2022, \mn@doi [\pasj] {10.1093/pasj/psab122}, 74, 247

\bibitem[\protect\citeauthoryear{Alabi et~al.,}{Alabi et~al.}{2016}]{Alabi_2016}
Alabi A.~B.,  et~al., 2016, \mn@doi [\mnras] {10.1093/mnras/stw1213}, 460, 3838

\bibitem[\protect\citeauthoryear{Alabi et~al.,}{Alabi et~al.}{2017}]{Alabi_2017}
Alabi A.~B.,  et~al., 2017, \mn@doi [\mnras] {10.1093/mnras/stx678}, 468, 3949

\bibitem[\protect\citeauthoryear{Alonso~Asensio, Dalla\, Bahé, Barnes  \& Kay}{Alonso~Asensio et~al.}{2020}]{Alonso_Asensio_2020}
Alonso~Asensio I.,  Dalla\ Vecchia C.,  Bahé Y.~M.,  Barnes D.~J.,   Kay S.~T.,  2020, \mn@doi [\mnras] {10.1093/mnras/staa861}, 494, 1859

\bibitem[\protect\citeauthoryear{Arena \& Bertin}{Arena \& Bertin}{2007}]{Arena_2007}
Arena S.~E.,  Bertin G.,  2007, \mn@doi [\aap] {10.1051/0004-6361:20066425}, 463, 921

\bibitem[\protect\citeauthoryear{{Arnaboldi} \& {Gerhard}}{{Arnaboldi} \& {Gerhard}}{2022}]{Arnaboldi_2022}
{Arnaboldi} M.,  {Gerhard} O.,  2022, \mn@doi [Front. Astron. Space Sci.] {10.3389/fspas.2022.872283}, 9, 403

\bibitem[\protect\citeauthoryear{{Arnaboldi}, {Freeman}, {Gerhard}, {Matthias}, {Kudritzki}, {M{\'e}ndez}, {Capaccioli}  \& {Ford}}{{Arnaboldi} et~al.}{1998}]{Arnaboldi_1998}
{Arnaboldi} M.,  {Freeman} K.~C.,  {Gerhard} O.,  {Matthias} M.,  {Kudritzki} R.~P.,  {M{\'e}ndez} R.~H.,  {Capaccioli} M.,   {Ford} H.,  1998, \mn@doi [\apj] {10.1086/306359}, 507, 759

\bibitem[\protect\citeauthoryear{Ascasibar \& Gottlöber}{Ascasibar \& Gottlöber}{2008}]{Ascasibar_2008}
Ascasibar Y.,  Gottlöber S.,  2008, \mn@doi [\mnras] {10.1111/j.1365-2966.2008.13160.x}, 386, 2022

\bibitem[\protect\citeauthoryear{Aubert, Pichon  \& Colombi}{Aubert et~al.}{2004}]{Aubert_2004}
Aubert D.,  Pichon C.,   Colombi S.,  2004, \mn@doi [\mnras] {10.1111/j.1365-2966.2004.07883.x}, 352, 376

\bibitem[\protect\citeauthoryear{Bahé et~al.,}{Bahé et~al.}{2017}]{Bahé_2017}
Bahé Y.~M.,  et~al., 2017, \mn@doi [\mnras] {10.1093/mnras/stx1403}, 470, 4186

\bibitem[\protect\citeauthoryear{Barbosa, Arnaboldi, Coccato, Gerhard, Mendes De~Oliveira, Hilker  \& Richtler}{Barbosa et~al.}{2018}]{Barbosa_2018}
Barbosa C.~E.,  Arnaboldi M.,  Coccato L.,  Gerhard O.,  Mendes De~Oliveira C.,  Hilker M.,   Richtler T.,  2018, \mn@doi [\aap] {10.1051/0004-6361/201731834}, 609, A78

\bibitem[\protect\citeauthoryear{Barnes \& Hut}{Barnes \& Hut}{1986}]{Barnes_Hut_1986}
Barnes J.,  Hut P.,  1986, \mn@doi [Nature] {10.1038/324446a0}, 324, 446

\bibitem[\protect\citeauthoryear{Bender, Kormendy, Cornell  \& Fisher}{Bender et~al.}{2015}]{Bender_2015}
Bender R.,  Kormendy J.,  Cornell M.~E.,   Fisher D.~B.,  2015, \mn@doi [\apj] {10.1088/0004-637X/807/1/56}, 807, 56

\bibitem[\protect\citeauthoryear{{Binney} \& {Tremaine}}{{Binney} \& {Tremaine}}{2008}]{Binney_Tremaine_2008}
{Binney} J.,  {Tremaine} S.,  2008, {Galactic Dynamics: Second Edition}.
{Princeton Univ. Press}, {Princeton}

\bibitem[\protect\citeauthoryear{Biviano et~al.,}{Biviano et~al.}{2013}]{Biviano_2013}
Biviano A.,  et~al., 2013, \mn@doi [\aap] {10.1051/0004-6361/201321955}, 558, A1

\bibitem[\protect\citeauthoryear{Blok}{Blok}{2010}]{Blok_2010}
Blok W. J. G.~d.,  2010, \mn@doi [Adv. Astron.] {10.1155/2010/789293}, 2010, 789293

\bibitem[\protect\citeauthoryear{Boardman et~al.,}{Boardman et~al.}{2017}]{Boardman_2017}
Boardman N.~F.,  et~al., 2017, \mn@doi [\mnras] {10.1093/mnras/stx1835}, 471, 4005

\bibitem[\protect\citeauthoryear{Borgani \& Guzzo}{Borgani \& Guzzo}{2001}]{Borgani_2001}
Borgani S.,  Guzzo L.,  2001, \mn@doi [Nature] {10.1038/409039A010.1038/35051000}, 409, 39

\bibitem[\protect\citeauthoryear{Brough et~al.,}{Brough et~al.}{2024}]{Brough_2024}
Brough S.,  et~al., 2024, \mn@doi [\mnras] {10.1093/mnras/stad3810}, 528, 771

\bibitem[\protect\citeauthoryear{Brown, Martin, Pearce, Hatch, Bahé  \& Dubois}{Brown et~al.}{2024}]{Brown_2024}
Brown H.~J.,  Martin G.,  Pearce F.~R.,  Hatch N.~A.,  Bahé Y.~M.,   Dubois Y.,  2024, \mn@doi [\mnras] {10.1093/mnras/stae2084}, 534, 431

\bibitem[\protect\citeauthoryear{Canepa, Brough, Lanusse, Montes  \& Hatch}{Canepa et~al.}{2025}]{Canepa_2025}
Canepa L.,  Brough S.,  Lanusse F.,  Montes M.,   Hatch N.,  2025, \mn@doi [\apj] {10.3847/1538-4357/adabc7}, 980, 245

\bibitem[\protect\citeauthoryear{Carter, Efstathiou, Ellis, Inglis  \& Godwin}{Carter et~al.}{1981}]{Carter_1981}
Carter D.,  Efstathiou G.,  Ellis R.~S.,  Inglis I.,   Godwin J.,  1981, \mn@doi [\mnras] {10.1093/mnras/195.1.15P}, 195, 15

\bibitem[\protect\citeauthoryear{Carter, Inglis, Ellis, Efstathiou  \& Godwin}{Carter et~al.}{1985}]{Carter_1985}
Carter D.,  Inglis I.,  Ellis R.~S.,  Efstathiou G.,   Godwin J.~G.,  1985, \mn@doi [\mnras] {10.1093/mnras/212.2.471}, 212, 471

\bibitem[\protect\citeauthoryear{Cañas, Elahi, Welker, Lagos, Power, Dubois  \& Pichon}{Cañas et~al.}{2019}]{Cañas_2019}
Cañas R.,  Elahi P.~J.,  Welker C.,  Lagos C. D.~P.,  Power C.,  Dubois Y.,   Pichon C.,  2019, \mn@doi [\mnras] {10.1093/mnras/sty2725}, 482, 2039

\bibitem[\protect\citeauthoryear{Cañas, Lagos, Elahi, Power, Welker, Dubois  \& Pichon}{Cañas et~al.}{2020}]{Cañas_2020}
Cañas R.,  Lagos C. d.~P.,  Elahi P.~J.,  Power C.,  Welker C.,  Dubois Y.,   Pichon C.,  2020, \mn@doi [\mnras] {10.1093/mnras/staa1027}, 494, 4314

\bibitem[\protect\citeauthoryear{{Chandrasekhar}}{{Chandrasekhar}}{1943}]{Chandrasekhar1943}
{Chandrasekhar} S.,  1943, \mn@doi [\apj] {10.1086/144517}, 97, 255

\bibitem[\protect\citeauthoryear{{Chen}, {Zu}, {Shao}  \& {Shan}}{{Chen} et~al.}{2022}]{Chen_2022}
{Chen} X.,  {Zu} Y.,  {Shao} Z.,   {Shan} H.,  2022, \mn@doi [\mnras] {10.1093/mnras/stac1456}, 514, 2692

\bibitem[\protect\citeauthoryear{Chun, Shin, Smith, Ko  \& Yoo}{Chun et~al.}{2023}]{Chun_2023}
Chun K.,  Shin J.,  Smith R.,  Ko J.,   Yoo J.,  2023, \mn@doi [\apj] {10.3847/1538-4357/aca890}, 943, 148

\bibitem[\protect\citeauthoryear{Chun, Shin, Ko, Smith  \& Yoo}{Chun et~al.}{2024}]{Chun_2024}
Chun K.,  Shin J.,  Ko J.,  Smith R.,   Yoo J.,  2024, \mn@doi [\apj] {10.3847/1538-4357/ad4a52}, 969, 142

\bibitem[\protect\citeauthoryear{Clowe, Gonzalez  \& Markevitch}{Clowe et~al.}{2004}]{Clowe_2004}
Clowe D.,  Gonzalez A.,   Markevitch M.,  2004, \mn@doi [\apj] {10.1086/381970}, 604, 596

\bibitem[\protect\citeauthoryear{Conroy, Wechsler  \& Kravtsov}{Conroy et~al.}{2007}]{Conroy_2007}
Conroy C.,  Wechsler R.~H.,   Kravtsov A.~V.,  2007, \mn@doi [\apj] {10.1086/521425}, 668, 826

\bibitem[\protect\citeauthoryear{Contini, De~Lucia, Villalobos  \& Borgani}{Contini et~al.}{2014}]{Contini_2014}
Contini E.,  De~Lucia G.,  Villalobos A.,   Borgani S.,  2014, \mn@doi [\mnras] {10.1093/mnras/stt2174}, 437, 3787

\bibitem[\protect\citeauthoryear{Contini, Yi  \& Kang}{Contini et~al.}{2019}]{Contini_2019}
Contini E.,  Yi S.~K.,   Kang X.,  2019, \mn@doi [\apj] {10.3847/1538-4357/aaf41f}, 871, 24

\bibitem[\protect\citeauthoryear{Contreras-Santos et~al.,}{Contreras-Santos et~al.}{2024}]{Contreras_Santos_2024}
Contreras-Santos A.,  et~al., 2024, \mn@doi [\aap] {10.1051/0004-6361/202348474}, 683, A59

\bibitem[\protect\citeauthoryear{Cui et~al.,}{Cui et~al.}{2014}]{Cui_2014}
Cui W.,  et~al., 2014, \mn@doi [\mnras] {10.1093/mnras/stt1940}, 437, 816

\bibitem[\protect\citeauthoryear{{Diego} et~al.,}{{Diego} et~al.}{2023}]{Diego2023}
{Diego} J.~M.,  et~al., 2023, \mn@doi [\aap] {10.1051/0004-6361/202345868}, 679, A159

\bibitem[\protect\citeauthoryear{{Diego} et~al.,}{{Diego} et~al.}{2024}]{Diego_2024}
{Diego} J.~M.,  et~al., 2024, \mn@doi [\aap] {10.1051/0004-6361/202349119}, 690, A114

\bibitem[\protect\citeauthoryear{Diemand, Moore  \& Stadel}{Diemand et~al.}{2004}]{Diemand_2004}
Diemand J.,  Moore B.,   Stadel J.,  2004, \mn@doi [\mnras] {10.1111/j.1365-2966.2004.07940.x}, 352, 535

\bibitem[\protect\citeauthoryear{Dolag, Murante  \& Borgani}{Dolag et~al.}{2010}]{Dolag_2010}
Dolag K.,  Murante G.,   Borgani S.,  2010, \mn@doi [\mnras] {10.1111/j.1365-2966.2010.16583.x}, 405, 1544

\bibitem[\protect\citeauthoryear{Dubois et~al.,}{Dubois et~al.}{2014}]{Dubois_2014}
Dubois Y.,  et~al., 2014, \mn@doi [\mnras] {10.1093/mnras/stu1227}, 444, 1453

\bibitem[\protect\citeauthoryear{Edwards et~al.,}{Edwards et~al.}{2020}]{Edwards_2020}
Edwards L. O.~V.,  et~al., 2020, \mn@doi [\mnras] {10.1093/mnras/stz2706}, 491, 2617

\bibitem[\protect\citeauthoryear{Elbert, Bullock, Garrison-Kimmel, Rocha, Oñorbe  \& Peter}{Elbert et~al.}{2015}]{Elbert_2015}
Elbert O.~D.,  Bullock J.~S.,  Garrison-Kimmel S.,  Rocha M.,  Oñorbe J.,   Peter A. H.~G.,  2015, \mn@doi [\mnras] {10.1093/mnras/stv1470}, 453, 29

\bibitem[\protect\citeauthoryear{Ettori, Donnarumma, Pointecouteau, Reiprich, Giodini, Lovisari  \& Schmidt}{Ettori et~al.}{2013}]{Ettori_2013}
Ettori S.,  Donnarumma A.,  Pointecouteau E.,  Reiprich T.~H.,  Giodini S.,  Lovisari L.,   Schmidt R.~W.,  2013, \mn@doi [Space Sci. Rev.] {10.1007/s11214-013-9976-7}, 177, 119

\bibitem[\protect\citeauthoryear{Eyles, Watt, Bertram, Church, Ponman, Skinner  \& Willmore}{Eyles et~al.}{1991}]{Eyles_1991}
Eyles C.~J.,  Watt M.~P.,  Bertram D.,  Church M.~J.,  Ponman T.~J.,  Skinner G.~K.,   Willmore A.~P.,  1991, \mn@doi [\apj] {10.1086/170251}, 376, 23

\bibitem[\protect\citeauthoryear{Feng}{Feng}{2010}]{Feng_2010}
Feng J.~L.,  2010, \mn@doi [ARA\&A] {https://doi.org/10.1146/annurev-astro-082708-101659}, 48, 495

\bibitem[\protect\citeauthoryear{Gifford, Miller  \& Kern}{Gifford et~al.}{2013}]{Gifford_2013}
Gifford D.,  Miller C.,   Kern N.,  2013, \mn@doi [\apj] {10.1088/0004-637X/773/2/116}, 773, 116

\bibitem[\protect\citeauthoryear{Golden-Marx et~al.,}{Golden-Marx et~al.}{2023}]{Golden-Marx_2023}
Golden-Marx J.~B.,  et~al., 2023, \mn@doi [\mnras] {10.1093/mnras/stad469}, 521, 478

\bibitem[\protect\citeauthoryear{Golden-Marx et~al.,}{Golden-Marx et~al.}{2025}]{Golden-Marx_2025}
Golden-Marx J.~B.,  et~al., 2025, \mn@doi [\mnras] {10.1093/mnras/staf277}, 538, 622

\bibitem[\protect\citeauthoryear{Gonzalez, Zaritsky  \& Zabludoff}{Gonzalez et~al.}{2007}]{Gonzalez_2007}
Gonzalez A.~H.,  Zaritsky D.,   Zabludoff A.~I.,  2007, \mn@doi [\apj] {10.1086/519729}, 666, 147

\bibitem[\protect\citeauthoryear{Gonzalez, Sivanandam, Zabludoff  \& Zaritsky}{Gonzalez et~al.}{2013}]{Gonzalez_2013}
Gonzalez A.~H.,  Sivanandam S.,  Zabludoff A.~I.,   Zaritsky D.,  2013, \mn@doi [\apj] {10.1088/0004-637X/778/1/14}, 778, 14

\bibitem[\protect\citeauthoryear{Grossman \& Narayan}{Grossman \& Narayan}{1989}]{Grossman_1989}
Grossman S.~A.,  Narayan R.,  1989, \mn@doi [\apj] {10.1086/167831}, 344, 637

\bibitem[\protect\citeauthoryear{{Haardt} \& {Madau}}{{Haardt} \& {Madau}}{1996}]{Haardt_1996}
{Haardt} F.,  {Madau} P.,  1996, \mn@doi [\apj] {10.1086/177035}, 461, 20

\bibitem[\protect\citeauthoryear{Haggar, Pearce, Gray, Knebe  \& Yepes}{Haggar et~al.}{2021}]{Haggar_2021}
Haggar R.,  Pearce F.~R.,  Gray M.~E.,  Knebe A.,   Yepes G.,  2021, \mn@doi [\mnras] {10.1093/mnras/stab064}, 502, 1191

\bibitem[\protect\citeauthoryear{Hartke, Arnaboldi, Longobardi, Gerhard, Freeman, Okamura  \& Nakata}{Hartke et~al.}{2017}]{Hartke_2017}
Hartke J.,  Arnaboldi M.,  Longobardi A.,  Gerhard O.,  Freeman K.~C.,  Okamura S.,   Nakata F.,  2017, \mn@doi [\aap] {10.1051/0004-6361/201730463}, 603, A104

\bibitem[\protect\citeauthoryear{Hatton, Devriendt, Ninin, Bouchet, Guiderdoni  \& Vibert}{Hatton et~al.}{2003}]{Hatton_2003}
Hatton S.,  Devriendt J. E.~G.,  Ninin S.,  Bouchet F.~R.,  Guiderdoni B.,   Vibert D.,  2003, \mn@doi [\mnras] {10.1046/j.1365-8711.2003.05589.x}, 343, 75

\bibitem[\protect\citeauthoryear{{He} et~al.,}{{He} et~al.}{2024}]{He_2024}
{He} J.,  et~al., 2024, \mn@doi [\apj] {10.3847/1538-4357/ad8882}, 976, 187

\bibitem[\protect\citeauthoryear{Hoekstra, Bartelmann, Dahle, Israel, Limousin  \& Meneghetti}{Hoekstra et~al.}{2013}]{Hoekstra_2013}
Hoekstra H.,  Bartelmann M.,  Dahle H.,  Israel H.,  Limousin M.,   Meneghetti M.,  2013, \mn@doi [Space Sci. Rev.] {10.1007/s11214-013-9978-5}, 177, 75

\bibitem[\protect\citeauthoryear{Hudson \& Robison}{Hudson \& Robison}{2018}]{Hudson_2018}
Hudson M.~J.,  Robison B.,  2018, \mn@doi [\mnras] {10.1093/mnras/sty844}, 477, 3869

\bibitem[\protect\citeauthoryear{{Ivezi{\'c}} et~al.,}{{Ivezi{\'c}} et~al.}{2019}]{Ivezic_2019}
{Ivezi{\'c}} {\v{Z}}.,  et~al., 2019, \mn@doi [\apj] {10.3847/1538-4357/ab042c}, 873, 111

\bibitem[\protect\citeauthoryear{Jiménez-Teja \& Dupke}{Jiménez-Teja \& Dupke}{2016}]{Jiménez-Teja_2016}
Jiménez-Teja Y.,  Dupke R.,  2016, \mn@doi [\apj] {10.3847/0004-637X/820/1/49}, 820, 49

\bibitem[\protect\citeauthoryear{Joshi, Parker, Wadsley  \& Keller}{Joshi et~al.}{2019}]{Joshi_2019}
Joshi G.~D.,  Parker L.~C.,  Wadsley J.,   Keller B.~W.,  2019, \mn@doi [\mnras] {10.1093/mnras/sty3119}, 483, 235

\bibitem[\protect\citeauthoryear{Kaviraj et~al.,}{Kaviraj et~al.}{2017}]{Kaviraj_2017}
Kaviraj S.,  et~al., 2017, \mn@doi [\mnras] {10.1093/mnras/stx126}, 467, 4739

\bibitem[\protect\citeauthoryear{Kelson, Zabludoff, Williams, Trager, Mulchaey  \& Bolte}{Kelson et~al.}{2002}]{Kelson_2002}
Kelson D.~D.,  Zabludoff A.~I.,  Williams K.~A.,  Trager S.~C.,  Mulchaey J.~S.,   Bolte M.,  2002, \mn@doi [\apj] {10.1086/341891}, 576, 720

\bibitem[\protect\citeauthoryear{Kimmig et~al.,}{Kimmig et~al.}{2025}]{Kimmig_2025}
Kimmig L.~C.,  et~al., 2025, \mn@doi [preprint (arXiv:2503.20857)] {10.48550/arXiv.2503.20857}

\bibitem[\protect\citeauthoryear{Kluge, Bender, Riffeser, Goessl, Hopp, Schmidt  \& Ries}{Kluge et~al.}{2021}]{Kluge_2021}
Kluge M.,  Bender R.,  Riffeser A.,  Goessl C.,  Hopp U.,  Schmidt M.,   Ries C.,  2021, \mn@doi [\apjs] {10.3847/1538-4365/abcda6}, 252, 27

\bibitem[\protect\citeauthoryear{Kluge et~al.,}{Kluge et~al.}{2024}]{Kluge_2024}
Kluge M.,  et~al., 2024, \mn@doi [preprint (arXiv:2405.13503)] {10.48550/arXiv.2405.13503}

\bibitem[\protect\citeauthoryear{{Knebe} et~al.,}{{Knebe} et~al.}{2013}]{Knebe2013}
{Knebe} A.,  et~al., 2013, \mn@doi [\mnras] {10.1093/mnras/stt1403}, 435, 1618

\bibitem[\protect\citeauthoryear{Kneib \& Natarajan}{Kneib \& Natarajan}{2011}]{Kneib_2011}
Kneib J.-P.,  Natarajan P.,  2011, \mn@doi [A\&AR] {10.1007/s00159-011-0047-3}, 19, 47

\bibitem[\protect\citeauthoryear{{Komatsu} et~al.,}{{Komatsu} et~al.}{2011}]{Komatsu_2011}
{Komatsu} E.,  et~al., 2011, \mn@doi [\apjs] {10.1088/0067-0049/192/2/18}, 192, 18

\bibitem[\protect\citeauthoryear{Lemze et~al.,}{Lemze et~al.}{2012}]{Lemze_2012}
Lemze D.,  et~al., 2012, \mn@doi [\apj] {10.1088/0004-637X/752/2/141}, 752, 141

\bibitem[\protect\citeauthoryear{Longobardi, Arnaboldi, Gerhard, Coccato, Okamura  \& Freeman}{Longobardi et~al.}{2013}]{Longobardi_2013}
Longobardi A.,  Arnaboldi M.,  Gerhard O.,  Coccato L.,  Okamura S.,   Freeman K.~C.,  2013, \mn@doi [\aap] {10.1051/0004-6361/201321652}, 558, A42

\bibitem[\protect\citeauthoryear{Loubser, Lagos, Babul, O’Sullivan, Jung, Olivares  \& Kolokythas}{Loubser et~al.}{2022}]{Loubser_2022}
Loubser S.~I.,  Lagos P.,  Babul A.,  O’Sullivan E.,  Jung S.~L.,  Olivares V.,   Kolokythas K.,  2022, \mn@doi [\mnras] {10.1093/mnras/stac1781}, 515, 1104

\bibitem[\protect\citeauthoryear{Lovell et~al.,}{Lovell et~al.}{2012}]{Lovell_2012}
Lovell M.~R.,  et~al., 2012, \mn@doi [\mnras] {10.1111/j.1365-2966.2011.20200.x}, 420, 2318

\bibitem[\protect\citeauthoryear{Lovell, Frenk, Eke, Jenkins, Gao  \& Theuns}{Lovell et~al.}{2014}]{Lovell_2014}
Lovell M.~R.,  Frenk C.~S.,  Eke V.~R.,  Jenkins A.,  Gao L.,   Theuns T.,  2014, \mn@doi [\mnras] {10.1093/mnras/stt2431}, 439, 300

\bibitem[\protect\citeauthoryear{Mamon, Cava, Biviano, Moretti, Poggianti  \& Bettoni}{Mamon et~al.}{2019}]{Mamon_2019}
Mamon G.~A.,  Cava A.,  Biviano A.,  Moretti A.,  Poggianti B.,   Bettoni D.,  2019, \mn@doi [\aap] {10.1051/0004-6361/201935081}, 631, A131

\bibitem[\protect\citeauthoryear{Markevitch, Gonzalez, Clowe, Vikhlinin, Forman, Jones, Murray  \& Tucker}{Markevitch et~al.}{2004}]{Markevitch_2004}
Markevitch M.,  Gonzalez A.~H.,  Clowe D.,  Vikhlinin A.,  Forman W.,  Jones C.,  Murray S.,   Tucker W.,  2004, \mn@doi [\apj] {10.1086/383178}, 606, 819

\bibitem[\protect\citeauthoryear{Martin, Pearce, Hatch, Contreras-Santos, Knebe  \& Cui}{Martin et~al.}{2024}]{Martin_2024}
Martin G.,  Pearce F.~R.,  Hatch N.~A.,  Contreras-Santos A.,  Knebe A.,   Cui W.,  2024, \mn@doi [\mnras] {10.1093/mnras/stae2488}, 535, 2375

\bibitem[\protect\citeauthoryear{Martínez-Lombilla et~al.,}{Martínez-Lombilla et~al.}{2023}]{Martínez-Lombilla_2023}
Martínez-Lombilla C.,  et~al., 2023, \mn@doi [\mnras] {10.1093/mnras/stac3119}, 518, 1195

\bibitem[\protect\citeauthoryear{{Mihos}}{{Mihos}}{2004}]{Mihos_2004}
{Mihos} J.~C.,  2004, in {Mulchaey} J.~S.,  {Dressler} A.,   {Oemler} A.,  eds, Clusters of Galaxies: Probes of Cosmological Structure and Galaxy Evolution. Cambridge University Press, Cambridge, p.~277

\bibitem[\protect\citeauthoryear{Montes}{Montes}{2022}]{Montes_2022}
Montes M.,  2022, \mn@doi [Nat. Astron.] {10.1038/s41550-022-01616-z}, 6, 308

\bibitem[\protect\citeauthoryear{Montes \& Trujillo}{Montes \& Trujillo}{2019}]{Montes_Trujillo_2019}
Montes M.,  Trujillo I.,  2019, \mn@doi [\mnras] {10.1093/mnras/sty2858}, 482, 2838

\bibitem[\protect\citeauthoryear{Montes, Brough, Owers  \& Santucci}{Montes et~al.}{2021}]{Montes_2021}
Montes M.,  Brough S.,  Owers M.~S.,   Santucci G.,  2021, \mn@doi [\apj] {10.3847/1538-4357/abddb6}, 910, 45

\bibitem[\protect\citeauthoryear{Moore}{Moore}{1994}]{Moore_1994}
Moore B.,  1994, \mn@doi [Nature] {10.1038/370629a0}, 370, 629

\bibitem[\protect\citeauthoryear{Muldrew, Pearce  \& Power}{Muldrew et~al.}{2011}]{Muldrew_2011}
Muldrew S.~I.,  Pearce F.~R.,   Power C.,  2011, \mn@doi [\mnras] {10.1111/j.1365-2966.2010.17636.x}, 410, 2617

\bibitem[\protect\citeauthoryear{Murante, Giovalli, Gerhard, Arnaboldi, Borgani  \& Dolag}{Murante et~al.}{2007}]{Murante_2007}
Murante G.,  Giovalli M.,  Gerhard O.,  Arnaboldi M.,  Borgani S.,   Dolag K.,  2007, \mn@doi [\mnras] {10.1111/j.1365-2966.2007.11568.x}, 377, 2

\bibitem[\protect\citeauthoryear{Ostriker \& Tremaine}{Ostriker \& Tremaine}{1975}]{Ostriker_1975}
Ostriker J.~P.,  Tremaine S.~D.,  1975, \mn@doi [\apj] {10.1086/181992}, 202, L113

\bibitem[\protect\citeauthoryear{Pillepich et~al.,}{Pillepich et~al.}{2014}]{Pillepich_2014}
Pillepich A.,  et~al., 2014, \mn@doi [\mnras] {10.1093/mnras/stu1408}, 444, 237

\bibitem[\protect\citeauthoryear{Pillepich et~al.,}{Pillepich et~al.}{2018}]{Pillepich_2018}
Pillepich A.,  et~al., 2018, \mn@doi [\mnras] {10.1093/mnras/stx3112}, 475, 648

\bibitem[\protect\citeauthoryear{{Planck Collaboration XIV,}}{{Planck Collaboration XIV,}}{2014}]{Planck_2014}
{Planck Collaboration XIV,} 2014, \mn@doi [\aap] {10.1051/0004-6361/201321591}, 571, A16

\bibitem[\protect\citeauthoryear{Pota et~al.,}{Pota et~al.}{2018}]{Pota_2018}
Pota V.,  et~al., 2018, \mn@doi [\mnras] {10.1093/mnras/sty2149}, 481, 1744

\bibitem[\protect\citeauthoryear{Proctor, Lagos, Ludlow  \& Robotham}{Proctor et~al.}{2023}]{Proctor_2023}
Proctor K.~L.,  Lagos C. d.~P.,  Ludlow A.~D.,   Robotham A. S.~G.,  2023, \mn@doi [\mnras] {10.1093/mnras/stad3341}, 527, 2624

\bibitem[\protect\citeauthoryear{Prunet, Pichon, Aubert, Pogosyan, Teyssier  \& Gottloeber}{Prunet et~al.}{2008}]{Prunet_2008}
Prunet S.,  Pichon C.,  Aubert D.,  Pogosyan D.,  Teyssier R.,   Gottloeber S.,  2008, \mn@doi [\apjs] {10.1086/590370}, 178, 179

\bibitem[\protect\citeauthoryear{Pulsoni et~al.,}{Pulsoni et~al.}{2018}]{Pulsoni_2018}
Pulsoni C.,  et~al., 2018, \mn@doi [\aap] {10.1051/0004-6361/201732473}, 618, A94

\bibitem[\protect\citeauthoryear{{Ramos-Almendares}, {Sales}, {Abadi}, {Doppel}, {Muriel}  \& {Peng}}{{Ramos-Almendares} et~al.}{2020}]{Ramos-Almendares2020}
{Ramos-Almendares} F.,  {Sales} L.~V.,  {Abadi} M.~G.,  {Doppel} J.~E.,  {Muriel} H.,   {Peng} E.~W.,  2020, \mn@doi [\mnras] {10.1093/mnras/staa551}, 493, 5357

\bibitem[\protect\citeauthoryear{Reina-Campos, Trujillo-Gomez, Deason, Kruijssen, Pfeffer, Crain, Bastian  \& Hughes}{Reina-Campos et~al.}{2022}]{Reina-Campos_2022}
Reina-Campos M.,  Trujillo-Gomez S.,  Deason A.~J.,  Kruijssen J. M.~D.,  Pfeffer J.~L.,  Crain R.~A.,  Bastian N.,   Hughes M.~E.,  2022, \mn@doi [\mnras] {10.1093/mnras/stac1126}, 513, 3925

\bibitem[\protect\citeauthoryear{Rudick, Mihos  \& McBride}{Rudick et~al.}{2006}]{Rudick_2006}
Rudick C.~S.,  Mihos J.~C.,   McBride C.,  2006, \mn@doi [\apj] {10.1086/506176}, 648, 936

\bibitem[\protect\citeauthoryear{Rudick, Christopher~Mihos, Frey  \& McBride}{Rudick et~al.}{2009}]{Rudick_2009}
Rudick C.~S.,  Christopher~Mihos J.,  Frey L.~H.,   McBride C.~K.,  2009, \mn@doi [\apj] {10.1088/0004-637X/699/2/1518}, 699, 1518

\bibitem[\protect\citeauthoryear{Scaramella et~al.,}{Scaramella et~al.}{2022}]{Scaramella_2022}
Scaramella R.,  et~al., 2022, \mn@doi [\aap] {10.1051/0004-6361/202141938}, 662, A112

\bibitem[\protect\citeauthoryear{Schuberth, Richtler, Hilker, Dirsch, Bassino, Romanowsky  \& Infante}{Schuberth et~al.}{2010}]{Schuberth_2010}
Schuberth Y.,  Richtler T.,  Hilker M.,  Dirsch B.,  Bassino L.~P.,  Romanowsky A.~J.,   Infante L.,  2010, \mn@doi [\aap] {10.1051/0004-6361/200912482}, 513, A52

\bibitem[\protect\citeauthoryear{Smith et~al.,}{Smith et~al.}{2016a}]{Smith_2016a}
Smith G.~P.,  et~al., 2016a, \mn@doi [\mnras] {10.1093/mnrasl/slv175}, 456, L74

\bibitem[\protect\citeauthoryear{Smith, Choi, Lee, Rhee, Sanchez-Janssen  \& Yi}{Smith et~al.}{2016b}]{Smith_2016b}
Smith R.,  Choi H.,  Lee J.,  Rhee J.,  Sanchez-Janssen R.,   Yi S.~K.,  2016b, \mn@doi [\apj] {10.3847/1538-4357/833/1/109}, 833, 109

\bibitem[\protect\citeauthoryear{Spergel \& Steinhardt}{Spergel \& Steinhardt}{2000}]{Spergel_2000}
Spergel D.~N.,  Steinhardt P.~J.,  2000, \mn@doi [Phys. Rev. Lett.] {10.1103/PhysRevLett.84.3760}, 84, 3760

\bibitem[\protect\citeauthoryear{Srisawat et~al.,}{Srisawat et~al.}{2013}]{Srisawat_2013}
Srisawat C.,  et~al., 2013, \mn@doi [\mnras] {10.1093/mnras/stt1545}, 436, 150

\bibitem[\protect\citeauthoryear{Strader et~al.,}{Strader et~al.}{2011}]{Strader_2011}
Strader J.,  et~al., 2011, \mn@doi [\apjs] {10.1088/0067-0049/197/2/33}, 197, 33

\bibitem[\protect\citeauthoryear{Sunyaev \& Zeldovich}{Sunyaev \& Zeldovich}{1972}]{Sunyaev_Zeldovich_1972}
Sunyaev R.~A.,  Zeldovich Y.~B.,  1972, Comment. Astrophys. Space Phys., 4, 173

\bibitem[\protect\citeauthoryear{{Sutherland} \& {Dopita}}{{Sutherland} \& {Dopita}}{1993}]{Sutherland_1993}
{Sutherland} R.~S.,  {Dopita} M.~A.,  1993, \mn@doi [\apjs] {10.1086/191823}, 88, 253

\bibitem[\protect\citeauthoryear{{Teyssier}}{{Teyssier}}{2002}]{Teyssier_2002}
{Teyssier} R.,  2002, \mn@doi [\aap] {10.1051/0004-6361:20011817}, 385, 337

\bibitem[\protect\citeauthoryear{{The Dark Energy Survey Collaboration,}}{{The Dark Energy Survey Collaboration,}}{2005}]{DES_2005}
{The Dark Energy Survey Collaboration,} 2005, \mn@doi [prepint (arXiv:astro-ph/0510346)] {10.48550/arXiv.astro-ph/0510346}

\bibitem[\protect\citeauthoryear{{Tormen}}{{Tormen}}{1997}]{Tormen1997}
{Tormen} G.,  1997, \mn@doi [\mnras] {10.1093/mnras/290.3.411}, 290, 411

\bibitem[\protect\citeauthoryear{Tulin \& Yu}{Tulin \& Yu}{2018}]{Tulin_2018}
Tulin S.,  Yu H.-B.,  2018, \mn@doi [Phy. Rep.] {10.1016/j.physrep.2017.11.004}, 730, 1

\bibitem[\protect\citeauthoryear{Tweed, Devriendt, Blaizot, Colombi  \& Slyz}{Tweed et~al.}{2009}]{Tweed_2009}
Tweed D.,  Devriendt J.,  Blaizot J.,  Colombi S.,   Slyz A.,  2009, \mn@doi [\aap] {10.1051/0004-6361/200911787}, 506, 647

\bibitem[\protect\citeauthoryear{{Villalobos}, {De Lucia}, {Borgani}  \& {Murante}}{{Villalobos} et~al.}{2012}]{Villalobos_2012}
{Villalobos} {\'A}.,  {De Lucia} G.,  {Borgani} S.,   {Murante} G.,  2012, \mn@doi [\mnras] {10.1111/j.1365-2966.2012.20667.x}, 424, 2401

\bibitem[\protect\citeauthoryear{Willman, Governato, Wadsley  \& Quinn}{Willman et~al.}{2004}]{Willman_2004}
Willman B.,  Governato F.,  Wadsley J.,   Quinn T.,  2004, \mn@doi [\mnras] {10.1111/j.1365-2966.2004.08312.x}, 355, 159

\bibitem[\protect\citeauthoryear{Wojtak, Łokas, Gottlöber  \& Mamon}{Wojtak et~al.}{2005}]{Wojtak_2005}
Wojtak R.,  Łokas E.~L.,  Gottlöber S.,   Mamon G.~A.,  2005, \mn@doi [\mnras] {10.1111/j.1745-3933.2005.00054.x}, 361, L1

\bibitem[\protect\citeauthoryear{Yoo et~al.,}{Yoo et~al.}{2024}]{Yoo_2024}
Yoo J.,  et~al., 2024, \mn@doi [\apj] {10.3847/1538-4357/ad2df8}, 965, 145

\bibitem[\protect\citeauthoryear{Zel’dovich}{Zel’dovich}{1970}]{Zeldovich_1970}
Zel’dovich Y.~B.,  1970, \aap, 5, 84

\bibitem[\protect\citeauthoryear{Zhang et~al.,}{Zhang et~al.}{2019}]{Zhang_2019}
Zhang Y.,  et~al., 2019, \mn@doi [\apj] {10.3847/1538-4357/ab0dfd}, 874, 165

\bibitem[\protect\citeauthoryear{{Zhang} et~al.,}{{Zhang} et~al.}{2024}]{Zhang_2024}
{Zhang} Y.,  et~al., 2024, \mn@doi [\mnras] {10.1093/mnras/stae1165}, 531, 510

\bibitem[\protect\citeauthoryear{{Zwicky}}{{Zwicky}}{1933}]{Zwicky_1933}
{Zwicky} F.,  1933, Helv. Phys. Acta, 6, 110

\bibitem[\protect\citeauthoryear{Zwicky}{Zwicky}{1951}]{Zwicky_1951}
Zwicky F.,  1951, \mn@doi [\pasp] {10.1086/126318}, 63, 61

\bibitem[\protect\citeauthoryear{van~den Bosch, Lewis, Lake  \& Stadel}{van~den Bosch et~al.}{1999}]{van_den_Bosch_1999}
van~den Bosch F.~C.,  Lewis G.~F.,  Lake G.,   Stadel J.,  1999, \mn@doi [\apj] {10.1086/307023}, 515, 50

\makeatother
\end{thebibliography}

\begin{appendix}

\section{Table of cluster properties}

We present key properties of the 12 {\sc Horizon-AGN} clusters in our sample in Table \ref{tab:cluster_properties}. All clusters are used in our $z\approx0$ analysis, but, due to breaks in their merger tree main progenitor branch, clusters Hrz078, Hrz137 and Hrz157 are excluded from the liberation time analysis presented in Figure \ref{fig:ser_liberation}. For the same reason, no $z_{50}$ is available for cluster Hrz137.

\begin{table*}
	\centering
	\caption{Properties of the 12 {\sc Horizon-AGN} clusters. $R_{178}$ is the radius of a sphere around the halo centre with average DM density $\bar{\rho}_\textrm{DM} = 178\,\rho_\textrm{M}$, $M_{178}$ is the total mass within $R_{178}$, and $M_{178,\star}$ is the total stellar mass within $R_{178}$. $z_{50}$ is the redshift at which half of the final mass is assembled into the cluster halo. No $z_{50}$ is available for cluster Hrz137 due to issues constructing a reliable merger tree. $f_\textrm{BCG}$, $f_\textrm{ICL}$ and $f_\textrm{sat}$ are the fractions of total cluster stellar mass in the BCG, ICL and satellite galaxies respectively, with a 100\,kpc spherical aperture used to separate the BCG and ICL. $\Delta v$ is the difference between the peaks of the DM and ICL velocity distributions, as described in Section\,\ref{sec:kinematics}.}
	\label{tab:cluster_properties}
	\begin{tabular}{ccccccccc}
		\hline
		$M_{\textrm{178}}$ ($\times 10^{14}\,\textrm{M}_{\odot}$) & $M_{\textrm{178,$\star$}}$ ($\times 10^{12}\,\textrm{M}_{\odot}$) & $R_{\textrm{178}}$ (Mpc) & $z_{\textrm{50}}$ & $f_\textrm{BCG}$ & $f_\textrm{ICL}$ & $f_\textrm{sat}$ & $\Delta v$ ($\textrm{km}\,\textrm{s}^{-1}$) & Cluster Name \\
		\hline
        1.18 & 2.84 & 1.13 & 1.24 & 0.35 & 0.29 & 0.36 & 270 & Hrz009 \\
        1.22 & 3.58 & 1.15 & 0.33 & 0.24 & 0.18 & 0.58 & 112 & Hrz183 \\
        1.23 & 3.28 & 1.15 & 0.75 & 0.24 & 0.24 & 0.52 & 172 & Hrz048 \\
        1.40 & 3.53 & 1.19 & 0.57 & 0.24 & 0.19 & 0.56 & 179 & Hrz046 \\
        1.46 & 4.09 & 1.21 & 0.27 & 0.24 & 0.17 & 0.59 & 445 & Hrz132 \\
        1.60 & 4.78 & 1.24 & 0.37 & 0.17 & 0.20 & 0.63 & 283 & Hrz071 \\
		1.70 & 4.73 & 1.29 & 0.42 & 0.12 & 0.22 & 0.65 & 227 & Hrz001 \\
        1.80 & 5.12 & 1.31 & 0.79 & 0.28 & 0.29 & 0.43 & 188 & Hrz157 \\
        2.32 & 6.74 & 1.42 & 0.47 & 0.20 & 0.28 & 0.52 & 216 & Hrz078 \\
        2.43 & 6.15 & 1.44 & 0.23 & 0.10 & 0.24 & 0.66 & 353 & Hrz174 \\
        2.66 & 6.43 & 1.46 & N/A & 0.12 & 0.21 & 0.67 & 376 & Hrz137 \\
        3.71 & 8.66 & 1.64 & 1.07 & 0.17 & 0.29 & 0.54 & 173 & Hrz049 \\
		\hline
	\end{tabular}
\end{table*}

\end{appendix}

\label{lastpage}
\end{document}